\documentstyle[emulateapj,apjfonts,psfig,amsmath,natbib,graphicx]{article}
\def\bdlong{2MASS J00361617+1821104}
\def\bd{2M0036+18}

\def\ociw{1}
\def\princeton{2}
\def\hubble{3}
\def\mcgill{4}
\def\stsci{5}
\def\ucsb{6}
\def\udel{7}
\def\steward{8}
\def\iac{9}
\def\ucf{10}
\def\ucb{11}
\def\toronto{12}
\def\rice{13}
\def\noao{14}
\def\uw{15}
\def\hamburg{16}

\begin{document}

\title{The Magnetic Properties of an L Dwarf Derived from
Simultaneous Radio, X-ray, and H$\alpha$ Observations}

\author{
E.~Berger\altaffilmark{\ociw,}\altaffilmark{\princeton,}\altaffilmark{\hubble},
R.~E.~Rutledge\altaffilmark{\mcgill},
I.~N.~Reid\altaffilmark{\stsci},
L.~Bildsten\altaffilmark{\ucsb},
J.~E.~Gizis\altaffilmark{\udel},
J.~Liebert\altaffilmark{\steward},
E.~Mart{\'{\i}}n\altaffilmark{\iac,}\altaffilmark{\ucf},
G.~Basri\altaffilmark{\ucb},
R.~Jayawardhana\altaffilmark{\toronto},
A.~Brandeker\altaffilmark{\toronto},
T.~A.~Fleming\altaffilmark{\steward},
C.~M.~Johns-Krull\altaffilmark{\rice},
M.~S.~Giampapa\altaffilmark{\noao},
S.~L.~Hawley\altaffilmark{\uw},
J.~H.~M.~M.~Schmitt\altaffilmark{\hamburg}
}

\altaffiltext{\ociw}{Observatories of the Carnegie Institution 
of Washington, 813 Santa Barbara Street, Pasadena, CA 91101}

\altaffiltext{\princeton}{Princeton University Observatory, 
Peyton Hall, Ivy Lane, Princeton, NJ 08544}

\altaffiltext{\hubble}{Hubble Fellow}

\altaffiltext{\mcgill}{Department of Physics, McGill University, 
Rutherford Physics Building, 3600 University Street, Montreal, 
QC H3A 2T8, Canada}

\altaffiltext{\stsci}{Space Telescope Science Institute, 
3700 San Martin Drive, Baltimore, MD 21218}

\altaffiltext{\ucsb}{Kavli Institute for Theoretical Physics, 
University of California at Santa Barbara, Kohn Hall, Santa 
Barbara, CA 93106}

\altaffiltext{\udel}{Department of Physics and Astronomy, 
University of Delaware, Newark, DE 19716}

\altaffiltext{\steward}{Department of Astronomy and Steward 
Observatory, University of Arizona, 933 North Cherry Avenue, 
Tucson, AZ 85721}

\altaffiltext{\iac}{Instituto de Astrof{\'{\i}}sica de Canarias, 
C/ V{\'{\i}}a L\'actea s/n, E-38200 La Laguna, Tenerife, Spain}

\altaffiltext{\ucf}{University of Central Florida, Department of
Physics, PO Box 162385, Orlando, FL 32816}

\altaffiltext{\ucb}{Astronomy Department, University of 
California, Berkeley, CA 94720}

\altaffiltext{\toronto}{Department of Astronomy and Astrophysics, 
University of Toronto, 60 St. George Street, Toronto, ON M5S 3H8, 
Canada}

\altaffiltext{\rice}{Department of Physics and Astronomy, Rice 
University, 6100 Main Street, MS-61 Houston, TX 77005}

\altaffiltext{\noao}{National Solar Observatory, National Optical 
Astronomy Observatories, Tucson, AZ 85726}

\altaffiltext{\uw}{University of Washington, Department of 
Astronomy, Box 351580, Seattle, WA 98195}

\altaffiltext{\hamburg}{Hamburger Sternwarte, Universitat 
Hamburg, Gojenbergsweg 112, 21029 Hamburg, Germany}

\begin{abstract} 
We present the first simultaneous, multi-wavelength observations of an
L dwarf, the L3.5 candidate brown dwarf \bdlong, conducted with the
Very Large Array, the {\it Chandra X-ray Observatory}, and the Kitt
Peak 4-m telescope.  We detect strongly variable and periodic radio
emission ($P=3$ hr) with a fraction of about $60\%$ circular
polarization.  No X-ray emission is detected to a limit of $L_X/
L_{\rm bol}\lesssim 2\times 10^{-5}$, several hundred times below the
saturation level observed in early M dwarfs.  Similarly, we do not
detect H$\alpha$ emission to a limit of $L_{H\alpha}/L_{\rm bol}
\lesssim 2\times 10^{-7}$, the deepest for any L dwarf observed to
date.  The ratio of radio to X-ray luminosity is at least four orders
of magnitude in excess of that observed in a wide range of active
stars (including M dwarfs) providing the first direct confirmation
that late-M and L dwarfs violate the radio/X-ray correlation.  The
radio emission is due to gyrosynchrotron radiation in a large-scale
magnetic field of about 175 G, which is maintained on timescales
longer than three years.  The detected 3-hour period may be due to (i)
the orbital motion of a companion at a separation of about five
stellar radii, similar to the configuration of RS CVn systems, (ii) an
equatorial rotation velocity of about $37$ km s$^{-1}$ and an
anchored, long-lived magnetic field, or (iii) periodic release of
magnetic stresses in the form of weak flares.  In the case of orbital
motion, the magnetic activity may be induced by the companion,
possibly explaining the unusual pattern of activity and the long-lived
signal.  We conclude that fully convective stars can maintain a
large-scale and stable magnetic field, but the lack of X-ray and
H$\alpha$ emission indicates that the atmospheric conditions are
markedly different than in early-type stars and even M dwarfs.
Similar observations are therefore invaluable for probing both the
internal and external structure of low mass stars and sub-stellar
objects, and for providing constraints on dynamo models.
\end{abstract}


\section{Introduction}
\label{sec:intro}

In recent years the stellar mass function has been extended at the
bottom of the main sequence and beyond with the discovery of numerous
very low mass stars and brown dwarfs (spectral types late-M, L, and T;
\citealt{mbd+97,dtf+99,mdb+99,rkl+99,lry+00,lgf+00,hck+02}).  Despite 
the significant observational and theoretical advances made since
their discovery (e.g., \citealt{bas00,cb00,bhl+01} and references
therein), many questions regarding the structure and formation of
these objects remain unanswered.  Not least among these is the
generation, amplification and dissipation of magnetic fields and their
influence on the coronae and chromospheres.  In turn, these provide a
window onto the physics of the internal convection, the structure of
the atmosphere, and its effect on the emergent radiation.  By analogy
with solar-type stars and M dwarfs, we expect that the magnetic
activity will be evident through coronal and chromospheric emission
(X-ray, H$\alpha$, and radio).

In the M dwarfs, observations in X-rays and H$\alpha$ reveal a
substantial fraction of active objects, peaking at $\sim 70\%$ around
spectral type M8 (e.g., \citealt{gmr+00,whw+04}).  Moreover, in both
bands the level of activity increases with rotation velocity, up to a
saturation level, at $v\sim 5$ km s$^{-1}$, of $L_X/L_{\rm bol}\approx
10^{-3}$ \citep{rgv85,tgs+93,pmm+03} and $L_{H\alpha}/L_{\rm bol}
\approx 10^{-3.5}$ \citep{mbs+02}.  These observations have been
interpreted in the context of coronal and chromospheric heating by
dissipation of magnetic fields, which in turn are powered by an
internal dynamo.  In stars earlier than M3 the dynamo is thought to be
powered by shearing motions at the radiative-convective transition
zone (the so-called $\alpha\Omega$ dynamo).  The increase in activity
with decreased mass and higher rotation velocity is understood as a
dependence of the dynamo on the Rossby number, $Ro=P/\tau_c$, where
$P$ is the rotation period and $\tau_c$ is the convective turnover
time (e.g., \citep{ssh+93}).  The saturation effect is still not fully
understood.

Surprisingly, the onset of full convection and the expected breakdown
of the $\alpha\Omega$ dynamo at spectral type M3 is not accompanied by
obvious changes in the level of H$\alpha$ and X-ray emission or the
rotation-activity relation.  This has been attributed to a growing
contribution from an $\alpha^2$ \citep{rwm+90} or a turbulent dynamo
\citep{ddr93} in which the shearing needed to generate and amplify the
fields comes from turbulent motion associated with the internal
convection.  However, beyond spectral type M7 there is a precipitous
drop in H$\alpha$ and X-ray persistent activity, and only a handful of
objects exhibit flares, $\sim 7\%$ in late-M dwarfs and $\sim 1\%$ in
L dwarfs (e.g., \citealt{rkg+99,gmr+00,rbm+00,lkc+03,whw+04} and see
Figure~\ref{fig:activity_sptype}).  Furthermore, the activity-rotation
relation no longer holds in late-M and L dwarfs with many rapid
rotators ($v\gtrsim 20$ km s$^{-1}$) exhibiting no discernible
activity \citep{bm95,mb03}.  The decrease in activity may be due to a
quenching of the turbulent dynamo and a transition to small-scale and
short-lived fields\footnotemark\footnotetext{Recent work by
\citet{dsb04} on magnetic field generation in convective rotating
stars suggests that large-scale fields can be generated and maintained
on relatively long timescales.  However, these authors use physical
conditions relevant for an M5 dwarf (and with an unusually high
surface temperature), while substantial changes in H$\alpha$ and X-ray
activity are manifested only for objects later than M7, and we are
interested in mid-L dwarfs here.}.  Moreover, the increasingly neutral
atmospheres of late-M and L dwarfs may hamper the dissipation of the
fields \citep{mbs+02}.  In effect, the increased rate of collisions
between charged and neutral particles may decouple the magnetic field
from the atmosphere, thereby suppressing the build-up of magnetic
stresses and the heating of the chromosphere and corona.

The above discussion suggests that radio emission, produced by
interaction of relativistic electrons with the magnetic field, should
also be suppressed in late-M and L dwarfs.  Observationally, the
empirical relation between the radio and X-ray luminosities of a wide
range of coronally-active stars, including many M dwarfs
\citep{gb93,bg94}, along with the few X-ray detections and upper
limits obtained to date, suggest an expected flux level below 0.1
$\mu$Jy (i.e., several hundred times lower than the detection
threshold of the Very Large Array).

However, radio observations of late-M and L dwarfs present a decidedly
different picture.  \citet{bbb+01} (hereafter, B01) detected radio
flares and persistent emission from the M9.5 brown dwarf LP944-20 four
orders of magnitude brighter than expected based on X-ray observations
\citep{rbm+00,mb02} and the radio/X-ray correlation.  A subsequent
survey of twelve additional objects (\citealt{ber02}; hereafter, B02)
revealed flares and persistent emission from three objects ranging
from M8.5 to L3.5.  In all cases the inferred magnetic fields
strengths were of the order of $100$ G.  In addition, of the three
detected objects, the two which have been observed in X-rays again
violated the radio/X-ray correlation by orders of magnitude.
Recently, \citet{pb03} detected radio emission from four out of seven
M7 to L4.5 dwarfs they observed with Australia Telescope Compact
Array, of which one exhibited a strong flare.  Thus, the incidence
rate of radio emission in late-M and L dwarfs may be as high as
$40\%$, much larger than in H$\alpha$ or X-rays.

Beyond the individual properties inferred from the unexpected radio
emission, two puzzling trends have emerged.  First, the fraction of
persistent radio emission compared to the bolometric luminosity
appears to increase with later spectral type, the exact opposite of
the trend observed in H$\alpha$ and X-ray activity.  Second, the radio
bright objects are those with high rotation velocities ($v{\rm
sin}i\gtrsim 20$ km s$^{-1}$), while significant non-detections are
associated with slow rotators ($v{\rm sin}i\lesssim 10$ km s$^{-1}$).
This implies that the rotation-activity relation holds in the radio,
despite being broken in H$\alpha$ and X-rays.

The glaring disparity between the trends observed in the various
activity indicators makes it clear that a complete picture of magnetic
activity in late-M and L dwarfs requires simultaneous,
multi-wavelength observations.  These observations can furthermore
constrain turbulent dynamo models, which despite significant progress
are still under-developed and under-constrained.  Here we present the
first such observations for any L dwarf, the L3.5 object \bdlong\
(hereafter, \bd).  In \S\ref{sec:obs} we provide the details of our
radio, X-ray, and H$\alpha$ observations.  We summarize the results in
\S\ref{sec:res} and address the radio emission mechanism and derive
the magnetic field properties in \S\ref{sec:emission}.  The violation
of the radio/X-ray correlation and its implications are discussed in
\S\ref{sec:gb}.  A possible periodicity observed in the radio emission
is assessed in \S\ref{sec:period} and implications for the magnetic
field generation mechanism are drawn in \S\ref{sec:mag}.  Finally, we
summarize the current trends in radio, H$\alpha$ and X-ray emission
from M and L dwarfs in \S\ref{sec:trends}.

\section{Observations}
\label{sec:obs}

We targeted the L3.5 dwarf \bd\ \citep{rkl+99} due to its relative
vicinity and previous observations of radio activity.  The object is
located at a distance of 8.8 pc \citep{dhv+02}, has a bolometric
luminosity, $L_{\rm bol}\approx 10^{-3.97}$ L$_\odot$
\citep{lgf+02,vhl+04}, a rotation velocity, $v{\rm sin}i\approx 15\pm
5$ km s$^{-1}$, measured from high-resolution optical spectra
\citep{sgh+01}, an inferred temperature of about $1900$ K
\citep{vhl+04}, an inferred radius of about $0.09\pm 0.01$ R$_\odot$
\citep{dhv+02}, a mass of about $0.076$ M$_\odot$ inferred from the
surface gravity, ${\rm log}g \approx 5.4$ \citep{sgh+01}, and an
inferred age of at least $1$ Gyr (e.g., \citealt{bhl+01}).  Thus, \bd\
is an object located near or below the hydrogen burning limit.  \bd\
has been detected in previous radio observations carried out in
September and October of 2001, with a strong $\sim 20$-min flare,
variable persistent emission, and a high fraction of circular
polarization (B02).

The observations presented here were carried out simultaneously in the
radio, X-ray, and optical on September 28, 2002.  Radio observations
commenced at 01:49 UT and ended at 09:39 UT (28.2 ks), X-ray
observations covered the range 01:10 to 07:17 UT (22 ks), and optical
spectroscopic observations covered the range 02:47 to 07:31 UT (17
ks).

\subsection{Radio}
\label{sec:rad}

Very Large Array (VLA\footnotemark\footnotetext{The VLA is operated by
the National Radio Astronomy Observatory, a facility of the National
Science Foundation operated under cooperative agreement by Associated
Universities, Inc.}) observations were conducted simultaneously at
4.86 and 8.46 GHz in the standard continuum mode with $2\times 50$ MHz
contiguous bands at each frequency.  We used fourteen antennas at 4.86
GHz and thirteen antennas at 8.46 GHz, with a staggered configuration
in each of the three array arms.  Eight-minute scans on \bd\ were
interleaved with 50 s scans on the phase calibrator J0042+233.  The
flux density scale was determined using the extragalactic source 3C 48
(J0137+331).

\begin{figure} 
\centerline{\psfig{file=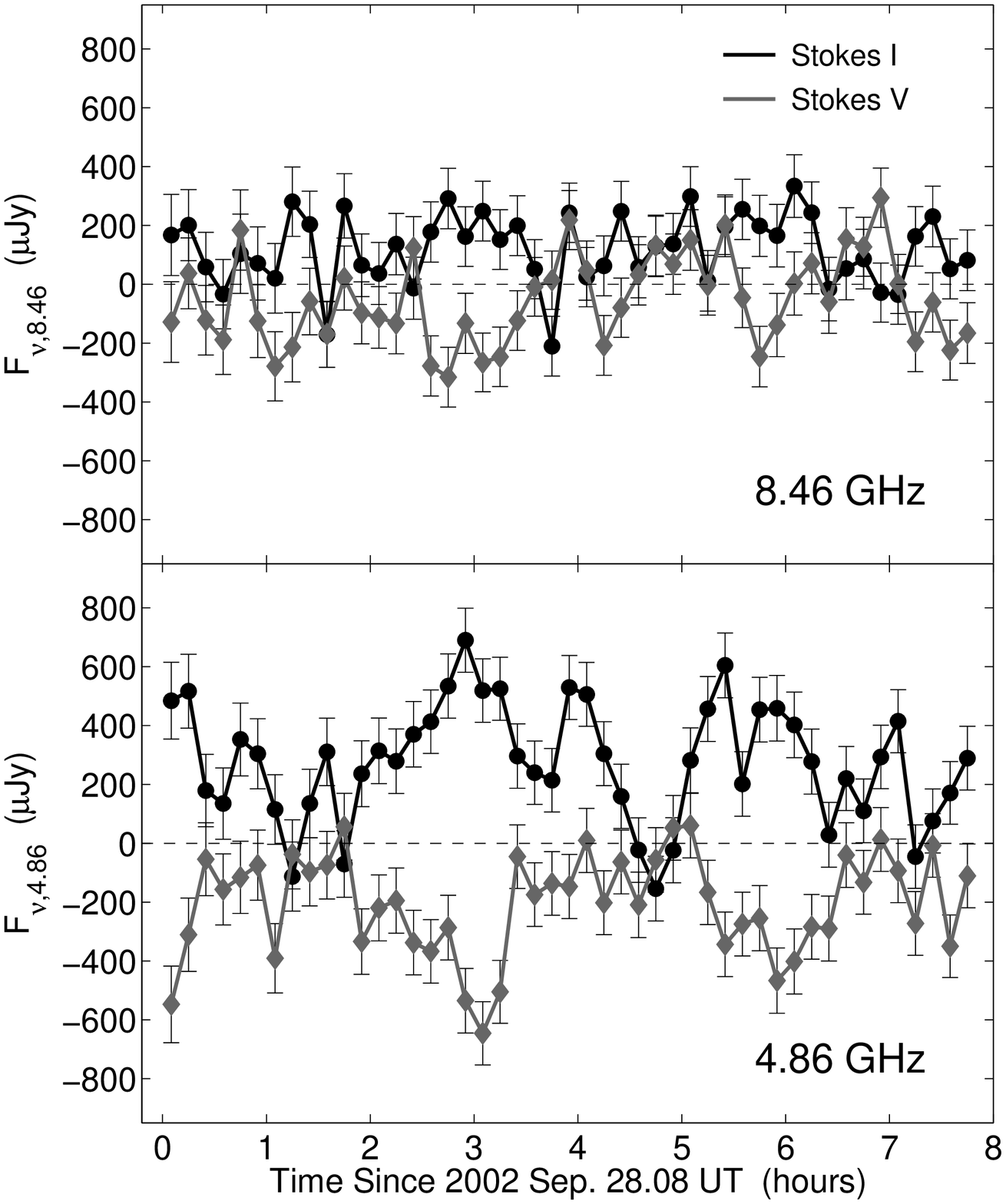,width=6.0in}}
\caption{Radio light curves of \bdlong\ at 8.5 GHz (top) and 4.9 GHz
(bottom).  Both total intensity (black circles) and circular
polarization (gray diamonds) are shown, with a 10-min time resolution.
The negative sign of the circularly-polarized flux indicates
left-handed polarization.  The light curves for both frequencies and
polarizations are variable, though with lower significance at 8.5 GHz
(\S\ref{sec:period}).
\label{fig:xc}}
\end{figure}

The data were reduced and analyzed using the Astronomical Image
Processing System (AIPS; Fomalont 1981).  The visibility data were
inspected for quality, and noisy points were removed.  To search for
flares, we constructed light curves using the following method.  We
removed all the bright field sources using the AIPS/IMAGR routine to
CLEAN the region around each source (with the exception of \bd), and
the AIPS/UVSUB routine to subtract the resulting source models from
the visibility data.  We then plotted the real part of the complex
visibilities at the position of \bd\ as a function of time using the
AIPS/UVPLT routine.  The subtraction of field sources is necessary
since their sidelobes and the change in the shape of the synthesized
beam during the observation result in flux variations over the map,
which may contaminate any real variability or generate false
variability.  We repeated this procedure for the background sources as
a check on the intrinsic noise properties of the maps.  The resulting
light curves are shown in Figures~\ref{fig:xc} and \ref{fig:c}.

We conducted follow-up observations with the VLA on 2005, Jan.~10 and
11 UT at 4.86 GHz.  A total of ten hours were obtained using the same
phase and flux calibrators as in the Sep.~2002 observation.  We
followed the same procedure outlined above to to produce light curves
(Figures~\ref{fig:0110} and \ref{fig:0111})

\subsection{X-rays}
\label{sec:xray}

The observations were made with the {\it Chandra}/ACIS-S3
(backside-illuminated chip).  \bd\ was offset from the on-axis focal
point by 4\arcmin\ to mitigate pileup in the event of a flare.  Data
were analyzed using CIAO v3.1.  We extracted data within a 4\arcsec\
circle centered on the source position, selecting counts in a low
energy band (0.2-3 keV) and a total energy band (0.2-10 keV).  Sources
were found using {\tt celldetect}.  To estimate the background we used
annuli centered at the source position, excluding other point sources
detected in the observation.  We adopt an energy conversion factor of
1 count = $4.5\times 10^{-12}$ erg cm$^{-2}$ s$^{-1}$ (0.2-8 keV),
consistent with a $kT$=1 keV Raymond-Smith plasma.
 
No X-ray source is detected at the optical position of \bd.  The
number of counts in the low and total energy bands were 3 and 5,
respectively, while we expected 2 and 4 counts from the background.
There is no evidence for flaring activity, or deviation from the
Poisson count rate distribution expected from a constant background.
For example, the shortest time-span which includes 3 counts is 6.4 ks.
For a detection of 5 counts, the 90\% confidence limit (found from an
average count rate of $<$9 counts per 21.5 ks) on the time-averaged
flux is $<1\times 10^{-15}$ erg cm$^{-2}$ s$^{-1}$.  We observed no
more than 1 count per 1 hr time period, corresponding to a 90\%
confidence limit on a 1 hr average peak flux ($<$3 counts/hr) of
$<4\times 10^{-15}$ erg cm$^{-2}$ s$^{-1}$.  At the distance of \bd\
the upper limit on persistent emission is $L_X\lesssim 9.3\times
10^{24}$ erg s$^{-1}$, or ${\rm log}(L_X/L_{\rm bol})\lesssim -4.65$.

\subsection{Optical}
\label{sec:opt}

\bd\ was monitored spectroscopically using the Multi-Aperture Red
Spectrometer (MARS) mounted on the Mayall 4-meter telescope at Kitt
Peak National Observatory.  A series of forty-eight 300-s exposures
covering $6400-10,100$\AA\ with a resolution of 3\AA\ pix$^{-1}$ were
obtained in clear and photometric conditions with reasonable seeing
($\sim $1.2\arcsec, although the initial observations were made at
airmass exceeding 2).  All of the spectra were flat-fielded,
extracted and wavelength calibrated using standard IRAF routines.
Observations of Feige 110 were used to set the flux scale.
                                                                               
\begin{figure} 
\centerline{\psfig{file=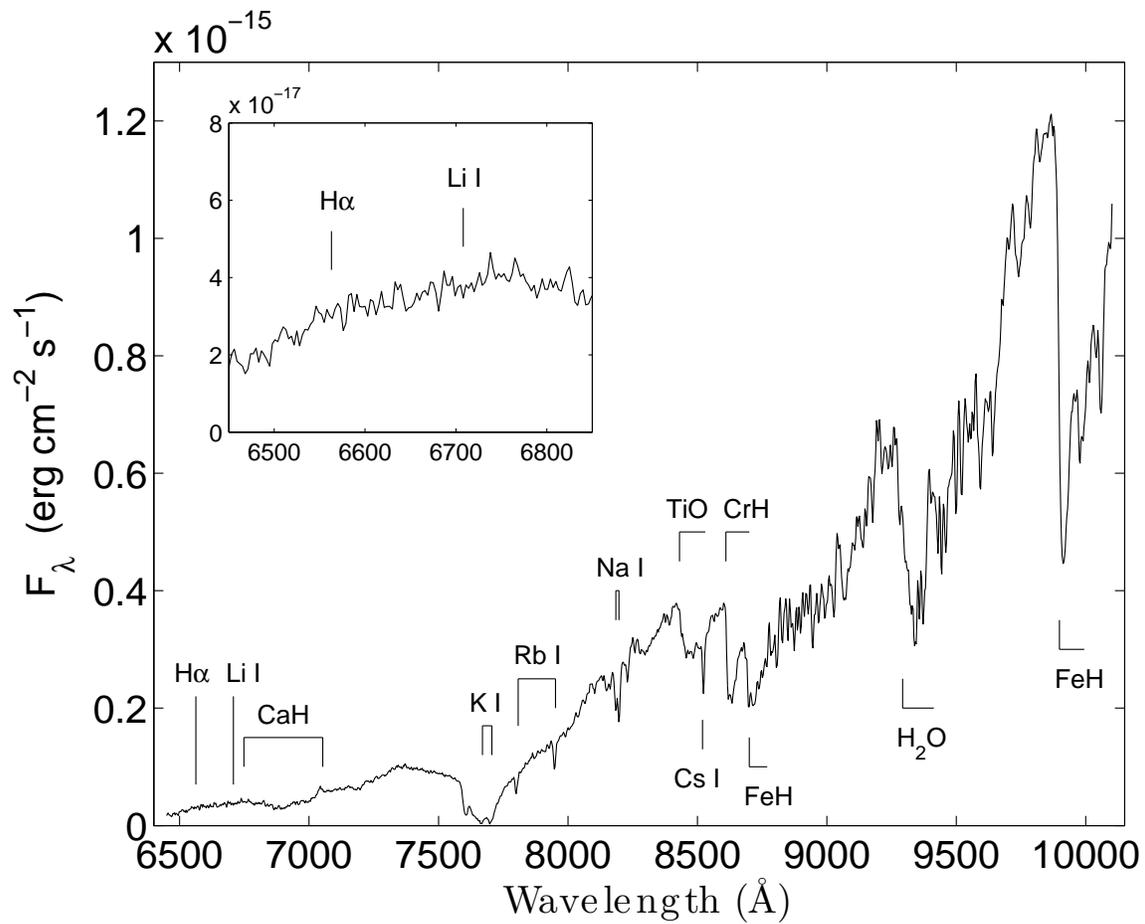,width=6.0in}}
\caption{Combined 4-hour optical spectrum of \bd.  The location of 
H$\alpha$ and prominent absorption features are marked.  Clearly, no
H$\alpha$ emission is detected above the continuum level (inset).
\label{fig:spec}}
\end{figure}

None of the spectra show evidence for H$\alpha$ emission.  The
individual 300-s spectra have signal-to-noise of about 10 to 15 at
these wavelengths, leading to an upper limit on the H$\alpha$
equivalent width of $\approx 1.5$\AA\ (or $F_{H\alpha}\lesssim 5\times
10^{-17}$ erg cm$^{-2}$ s$^{-1}$).  Combining all the individual
spectra (Figure~\ref{fig:spec}), we find no evidence for emission
exceeding an equivalent width of 0.4\AA, or $F_{H\alpha}\lesssim
1\times 10^{-17}$ erg cm$^{-2}$ s$^{-1}$.  At the distance of \bd\
this translates to $L_{H\alpha}\lesssim 9.3\times 10^{22}$ erg
s$^{-1}$, or ${\rm log}(L_{H\alpha}/L_{\rm bol})\lesssim -6.65$.

Separately, on 2004 December 5 we obtained two 10-min high-resolution
spectra with the Magellan Inamori Kyocera Echelle (MIKE) Spectrograph,
separated by 103 min.  The purpose of this observation was to measure
a possible radial velocity induced by a close-in companion (see
\S\ref{sec:period}).  A 0.35\arcsec\ slit was used with $2\times 2$
binning, providing a spectral resolution of about 27,000 in the red or
about $0.11$\AA\ per pixel.  The spectra were flat-fielded, rectified,
wavelength calibrated and extracted using a MIKE data reduction
pipeline.  The rms wavelength residuals were of the order of 0.005\AA\
and the signal-to-noise in the relevant echelle orders was
approximately ten and six in the first and second spectrum,
respectively.

\section{Basic Properties of the Radio Emission}
\label{sec:res}

While neither X-ray nor H$\alpha$ emission were detected from \bd,
radio emission was detected with an average flux of $259\pm 19$
$\mu$Jy and $134\pm 16$ $\mu$Jy at 4.9 and 8.5 GHz, respectively
(Figure~\ref{fig:xc}).  The resulting spectral index ($F_\nu\propto
\nu^\alpha$) is $\alpha=-1.2\pm 0.3$, typical of radio emission
observed from M dwarfs \citep{gsb+93}.  The emission is strongly
circularly polarized with an average of $f_c\approx -73\pm 8\%$ at 4.9
GHz and $f_c\approx -60\pm 15\%$ at 8.5 GHz.  The negative sign
indicates that the polarization direction is left-handed.  We place a
limit of $\lesssim 20\%$ ($3\sigma$) on the fraction of linear
polarization at 4.9 GHz and $\lesssim 35\%$ at 8.5 GHz.  In the
follow-up observations we detected emission from \bd\ at a level of
$152\pm 9$ $\mu$Jy, with a fraction of circular polarization,
$f_c\approx -46\pm 7\%$.

For comparison, two past observations of \bd\ at 8.5 GHz (B02)
revealed a similar flux level, $135$ and $330$ $\mu$Jy, but a lower
fraction of circular polarization, $-35\%$ and $-13\%$.  One of the
two observations also uncovered a 20-min flare with a peak flux of
about $720$ $\mu$Jy and a fraction of circular polarization,
$f_c\approx -62\%$.  No such flares are observed here suggesting that
their rate of occurrence, taking into account all three observations,
is $<0.04$ hr$^{-1}$.  The flare duty cycle, defined as the flare
duration relative to the total observing time, is $\lesssim 1.5\%$.

The brightness temperature of the radio emission, an important 
discriminant of the emission mechanism, is given by
\begin{equation} 
T_b\approx 2\times 10^9 F_{\rm \nu,mJy}\nu^{-2}_{\rm GHz} 
d^2_{\rm pc} (R/R_J)^{-2}\,\,{\rm K},  
\end{equation}
where $R_s/R_J\approx 1$ is the radius in units of Jupiter radii.  We
find that $T_b\approx 2\times 10^8$ K if the emission is produced in
the corona at a radius of $R\sim 3R_s\sim 3R_J$
\citep{lg83,bhl89,lpl+00} and with a covering fraction of $100\%$.  On
the other hand, if the emission is confined to a significantly smaller
region (e.g., a coronal loop with $R\lesssim 0.1R_s$) then $T_b$ may
exceed $10^{11}$ K.  In the case of gyromagnetic radiation (see
\S\ref{sec:gyro}), the inverse Compton catastrophe limit, $T_b\lesssim
10^{12}$ K, defines a minimum size for the emission region of about
$0.08R_s\approx 5.0\times 10^8$ cm.

\section{Emission Mechanism} 
\label{sec:emission}

Several radiation mechanisms can give rise to the observed radio
emission (e.g., \citealt{gue02}): Bremsstrahlung radiation,
gyromagnetic radiation, and coherent processes (plasma radiation or
electron cyclotron maser).  We can rule out bremsstrahlung radiation
in this case since the expected spectrum is flat in the optically thin
regime, whereas here $\alpha\approx -1.2$.

Coherent emission processes, which have been invoked in cases of high
brightness temperature ($T_b>10^{12}$ K) solar and stellar flares with
strong circular polarization, are also unlikely in this case.  This is
primarily because the emission is expected to have a narrow bandwidth
around the fundamental or second harmonic of the plasma frequency,
$\nu_p\approx 9000n_e^{1/2}$ Hz, or the cyclotron frequency,
$\nu_c\approx 2.8\times 10^6B$ Hz, whereas here emission is detected
at both 4.9 and 8.5 GHz.  In addition, the typical emission timescale
of coherent radiation flares is much shorter than the 7.8-hour
duration of the radio emission detected in this case.

\subsection{Gyromagnetic Radiation}
\label{sec:gyro}

Gyromagnetic radiation can arise from a thermal or non-thermal
distribution of electrons, with the latter typically assumed to follow
a power law, $n(\gamma)\propto \gamma^{-p}$, above some cutoff,
$\gamma_m$.  The emission properties further depend on the harmonic
number, $s$, or equivalently the typical energy of the electrons
$\gamma=s^{1/3}$.  Cyclotron emission is produced when $s\!\lesssim\!
10$ (non-relativistic electrons, typically thermal distribution),
gyrosynchrotron when $s\!\approx\! 10-100$ (mildly relativistic
electrons), and synchrotron when $s\!\gtrsim\! 100$ (relativistic
electrons).  The frequency at which the bulk of the radiation is
emitted is given by $\nu_m\approx 2.8\times 10^6sB$ Hz, where $B$ is
the magnetic field strength.

Cyclotron radiation is unlikely in this case since the emission is
expected to be concentrated in narrow bands at the harmonic
frequencies, as opposed to the detected signal at 4.9 and 8.5 GHz.
Similarly, gyrosynchrotron emission from a thermal plasma is unlikely
since the expected optically thin spectral index, $\alpha \approx -8$,
is significantly steeper than the observed value.  Finally,
synchrotron emission is unlikely because the expected degree of
circular polarization is strongly suppressed compared to the value of
about $70\%$ measured here.  Therefore, the most likely emission
mechanism in the case of \bd\ is gyrosynchrotron radiation from a
non-thermal distribution of electrons.

The emission properties in the gyrosynchrotron case depend both on the
fundamental source parameters -- magnetic field $B$, electron density
$n_e$, and line-of-sight source scale $L$ -- and the angle between the
line of sight and the magnetic field, $\theta$ \citep{dm82}.  Since
the latter cannot be inferred from the present data we take
$\theta=\pi/3$ as a representative value.  The spectral index in the
optically thin case, $\alpha\approx 1.2-0.9p$, indicates that
$p\approx 2.7$.  This value is similar to those found in several early
M dwarfs \citep{gsb+93}.  Given the value of $p$, the fraction of
circular polarization at 4.9 GHz is given by $f_c\approx 0.025
B^{0.53}\approx 0.6$, from which we infer $B\approx 400$ G.  For the
full range of possible $\theta$ values the magnetic field strength
ranges from about 80 to $1.9\times 10^3$ G.  These values indicate
that the harmonic number is $s<22$, with the gyrosynchrotron limit,
$s>10$, corresponding to $B<175$ G, or $\theta<\pi/4$.  Below we use
the latter value for the magnetic field strength.

The frequency at which the gyrosynchrotron spectrum peaks is given by
$\nu_m\approx 6.1\times 10^5(n_eL)^{0.24}<4.86\times 10^9$ Hz, while
the flux in the optically thin regime is $F_\nu\approx 1.1\times
10^{-34}n_eL R^2\approx 205$ $\mu$Jy, using the average flux at 4.9
GHz.  Solving for the unknown column density and radius of the
emission region we find $n_eL<1.8\times 10^{16}$ cm$^{-2}$ and
$R\approx 1\times 10^{10}$ cm, or about $1.4R_s$.  The resulting
brightness temperature is $T_b\approx 7\times 10^8$, consistent with
the interpretation of the radio emission as gyrosynchrotron radiation.
Assuming that $L=R$, the electron density is $n_e\approx 1.6\times
10^6$ cm$^{-3}$.

Several conclusions can be drawn from this discussion.  First, the
magnetic field strength is larger than the average solar field or the
magnetic field of Jupiter, but is somewhat lower than the $\sim 1$ kG
fields inferred in some active early M dwarfs
\citep{sdj+75,sl85,hsr91,jv96,skz+01}.  We note however, that the 
surface magnetic field of \bd\ may be an order of magnitude larger
than that inferred for the emission region if $L\approx R\approx
1.4R_s$, placing it in the same range of early M dwarfs.  Second, the
radio emission, and in particular the high fraction of circular
polarization, requires a large-scale, long-lived ordered magnetic
field with a large covering fraction.  This is again similar to the
conditions inferred for early M dwarfs \citep{sl85,jv96}.  Finally,
the density of emitting electrons is similar to that inferred in the
emission zones (coronae) of early M dwarfs \citep{skz+01,oha+04}.
Thus, the physical conditions in the L3.5 \bd\ resemble those found in
active early M dwarf, but the H$\alpha$ and X-ray emission are
strongly suppressed.

We finally note that the plasma-beta, $\beta=2\mu_0P/B^2$, of the
corona, using $n\sim 1.6\times 10^6$ cm$^{-3}$, a typical temperature
$T\sim 2\times 10^6$ K \citep{grk+96}, and $B\sim 175$ G, is
$\beta\sim 3\times 10^{-8}$.  The typical photospheric pressure of an
L dwarf is $P\sim 3\times 10^5$ dyne cm$^{-2}$ \citep{bhl+01},
resulting in $\beta\sim 2$. These conditions are again similar to
those found in M dwarfs \citep{grk+96} and the Sun.  Thus, even small
perturbations in the coronal field of \bd\ are capable of supporting
the pressure gradients produced by the radio-emitting material.

\section{Violation of the Radio/X-ray Correlation}
\label{sec:gb}

\begin{figure} 
\centerline{\psfig{file=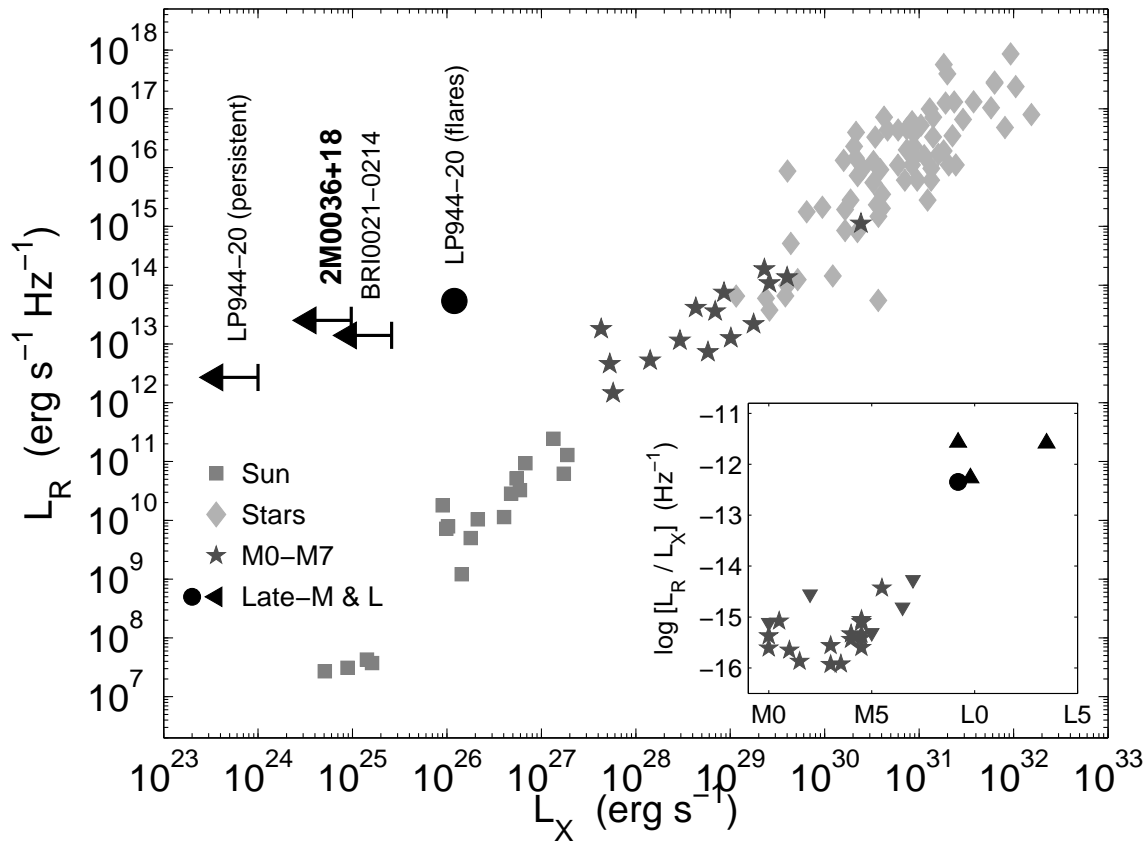,width=6.0in}}
\caption{Radio versus X-ray luminosity for stars exhibiting coronal 
activity.  Data for late-M and L dwarfs are from B01 and B02, while
other data are taken from \citet{gue02} and references therein.  Data
points for the Sun include impulsive and gradual flares, as well as
microflares.  The strong correlation between $L_R$ and $L_X$ is
evident and extends to spectral type M7 (see inset).  Clearly,
\bdlong\ and the other late-M and L dwarfs detected in the radio to
date violate the correlation by many orders of magnitude.
\label{fig:gb}}
\end{figure}

The simultaneous radio and X-ray observations allow us for the first
time to directly investigate the radio/X-ray correlation in an L
dwarf.  Similar observations of coronally active stars (including the
Sun) reveal a tight correlation between $L_R$ and $L_X$
(Figure~\ref{fig:gb}; \citealt{gb93,bg94}).  This correlation holds
for both flaring and persistent emission, and for single stars and
binary systems.  For M dwarfs the correlation holds to spectral type
M7 \citep{gsb+93}.  The persistent emission follows $L_R\approx
3\times 10^{-16}L_X$ Hz$^{-1}$ \citep{gb93}, and extends over six
orders of magnitude in $L_R$.  The relation for radio and X-ray flares
is non-linear, $L_R\propto 1.3\times 10^{-26}L_X^{0.73}$ Hz$^{-1}$
\citep{bg94}, and holds over eight orders of magnitude in $L_R$.

For \bd\ we find $L_R\approx 2.5\times 10^{13}$ erg s$^{-1}$ Hz$^{-1}$
and $L_X<9.7\times 10^{24}$ erg s$^{-1}$.  Thus, $L_R/L_X\gtrsim
2.6\times 10^{-12}$ Hz$^{-1}$ is at least 8600 times higher than
predicted by the radio/X-ray correlation (Figure~\ref{fig:gb}).  A
similar result, from non-simultaneous data, was obtained for the M9.5
brown dwarf LP944-20 (B01) with a ratio that exceeded the expected
value by a factor of $10^4$ (flares) and $>8500$ (persistent
emission).  Similarly, for the M9.5 dwarf BRI\,0021-0214 the radio
luminosity exceeded the X-ray upper limit by at least a factor of
$10^3$ (B02).  Thus, the radio and X-ray luminosities appear to
de-correlate over a narrow range in spectral type (between about M7
and M9) and remain uncorrelated at least to mid-L
(Figure~\ref{fig:gb}).  Incidentally, this is the same range over
which both the fraction of H$\alpha$ and X-ray active sources and the
strength of the activity drop precipitously.

The violation of the radio/X-ray correlation is surprising given that
the magnetic and emission region properties derived for \bd\ are not
dissimilar from those found on early M dwarfs (\S\ref{sec:gyro}).
This suggests that either the origin of the magnetic field is
different than in M dwarfs or the atmospheric conditions are not
conducive for the support of a chromosphere and/or a corona.

The radio/X-ray correlation has been interpreted in the context of
magnetic reconnection.  In this scenario, the flares are produced when
coronal magnetic loops reconnect, release energy, and create a current
sheet along which ambient electrons are accelerated.  The accelerated
electrons drive an outflow of hot plasma into the corona as they
interact with and heat the underlying chromospheric material.  The
interaction of the outflowing plasma with the electrons produces X-ray
emission via the bremsstrahlung process \citep{neu68,hfs+95,gbs+96}.
This so-called Neupert effect points to a causal connection between
particle acceleration, which is the source of radio emission, and
plasma heating, which results in X-ray emission.  Thus, the X-ray
thermal energy should simply be related by a constant of
proportionality to the integrated radio flux.

\citet{gb93} discuss this correlation in terms of the relation
$L_R/L_X\approx 10^{-22}B^{2.5}(a/b)\tau_0(1+\alpha)$ Hz$^{-1}$, where
$a$ is the fractional efficiency of accelerating electrons, $b$ is the
fraction of coronal energy radiated in X-rays, $\tau_0$ is the average
lifetime of the low-energy electrons, and $\alpha$ is the power law
relating the electron lifetime to its energy.  The observed tight
correlation suggests that the combination $B^{2.5}(a/b)\tau_0$ is
nearly unchanged for a wide range of stars.  Since in the case of \bd\
and the other radio active late-M and L dwarfs $L_R/L_X$ is higher by
several orders of magnitude while the inferred value of $B$ is not
unusual compared to M dwarfs, this indicates that either the fraction
of energy emitted in X-rays, $b$, is suppressed by a factor of $\sim
10^4$, or the typical lifetime of the electrons, $\tau_0$, is longer
by a similar factor.  In particular, for $a/b$ of order unity and
$B\approx 175$ G, we find a mean electron lifetime $\tau\sim 1$ d.  In
this case, the X-ray emission is suppressed because the electrons do
not lose their energy to coronal heating on a sufficiently rapid
timescale.

On the other hand, the violation of the radio/X-ray correlation may be
due to an inefficient production of X-rays, i.e., a low value of $b$.
In the case of L dwarfs this may be explained by the increasingly
neutral atmospheres and the reduction in the density of ions available
for X-ray emission \citep{mbs+02}.  However, it is not clear if this
can account for the late-M dwarfs LP944-20 and BRI\,0021-0214, or for
the sudden transition between spectral types M7 and M9.  More likely,
the reduction in X-ray luminosity, or the over-production of radio
emission, may be related to the magnetic structure across the
atmosphere, chromosphere and corona or efficient trapping of the
electrons (i.e., long lifetime).  We note that recent radio and X-ray
observations of the dM4.5e star EV Lac also reveal a breakdown of the
Neupert effect, which \citet{oha+04} interpret in the same context
discussed here, namely efficient trapping or a low value of $b$.

\section{Variability and Periodicity}
\label{sec:period}

\begin{figure} 
\centerline{\psfig{file=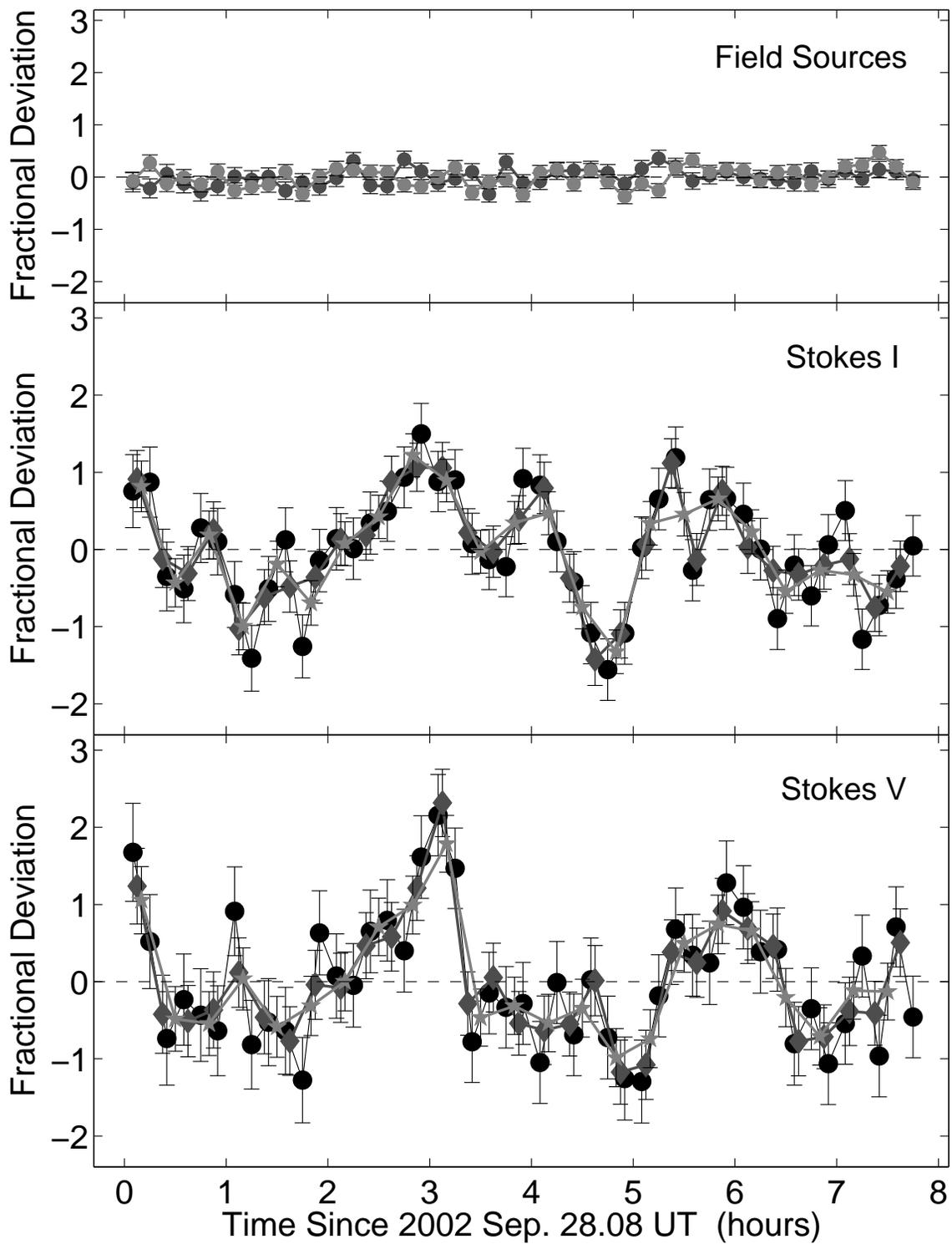,width=6.0in}}
\caption{Fractional deviation relative to the mean of the circular 
polarization (bottom) and total intensity (middle) light curves of
\bdlong\ at 4.9 GHz.  Time resolutions of 10, 15, and 20 minutes are
plotted.  The top panel shows the light curves of two field sources
with a 10-min time resolution.  Clearly, the emission from
\bd\ is highly variable independent of time resolution.  The
light curve of \bd\ also appears to be periodic with a period of about
3 hours.
\label{fig:c}}
\end{figure}

\begin{figure} 
\centerline{\psfig{file=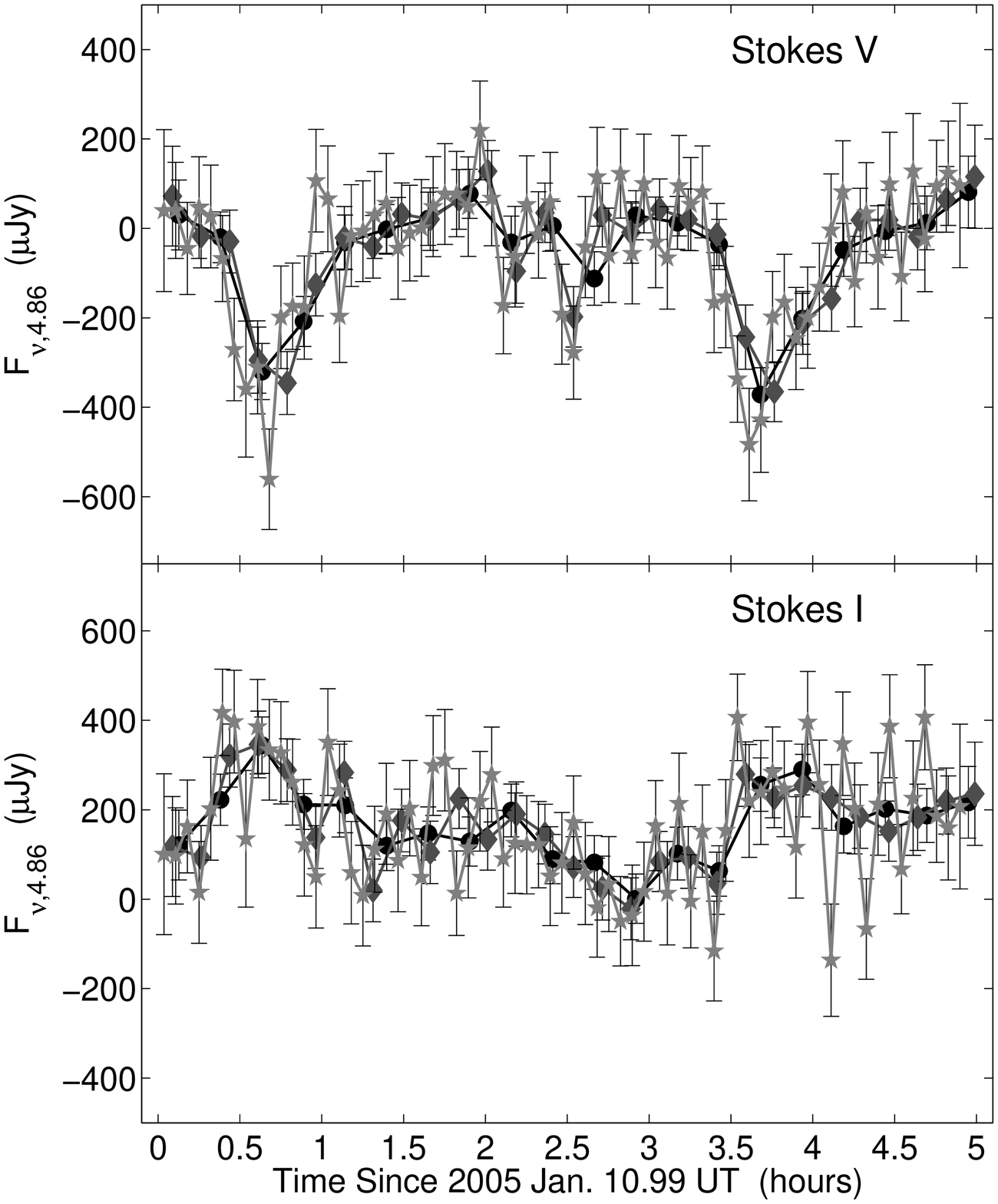,width=6.0in}}
\caption{Follow-up observations of \bd\ at 4.86 GHz taken on 2005, 
Jan.~10.99 UT for a total of 5 hours.  The 3-hour period detected in
the initial observations from Sep.~2002 is clearly seen.
\label{fig:0110}}
\end{figure}

\begin{figure} 
\centerline{\psfig{file=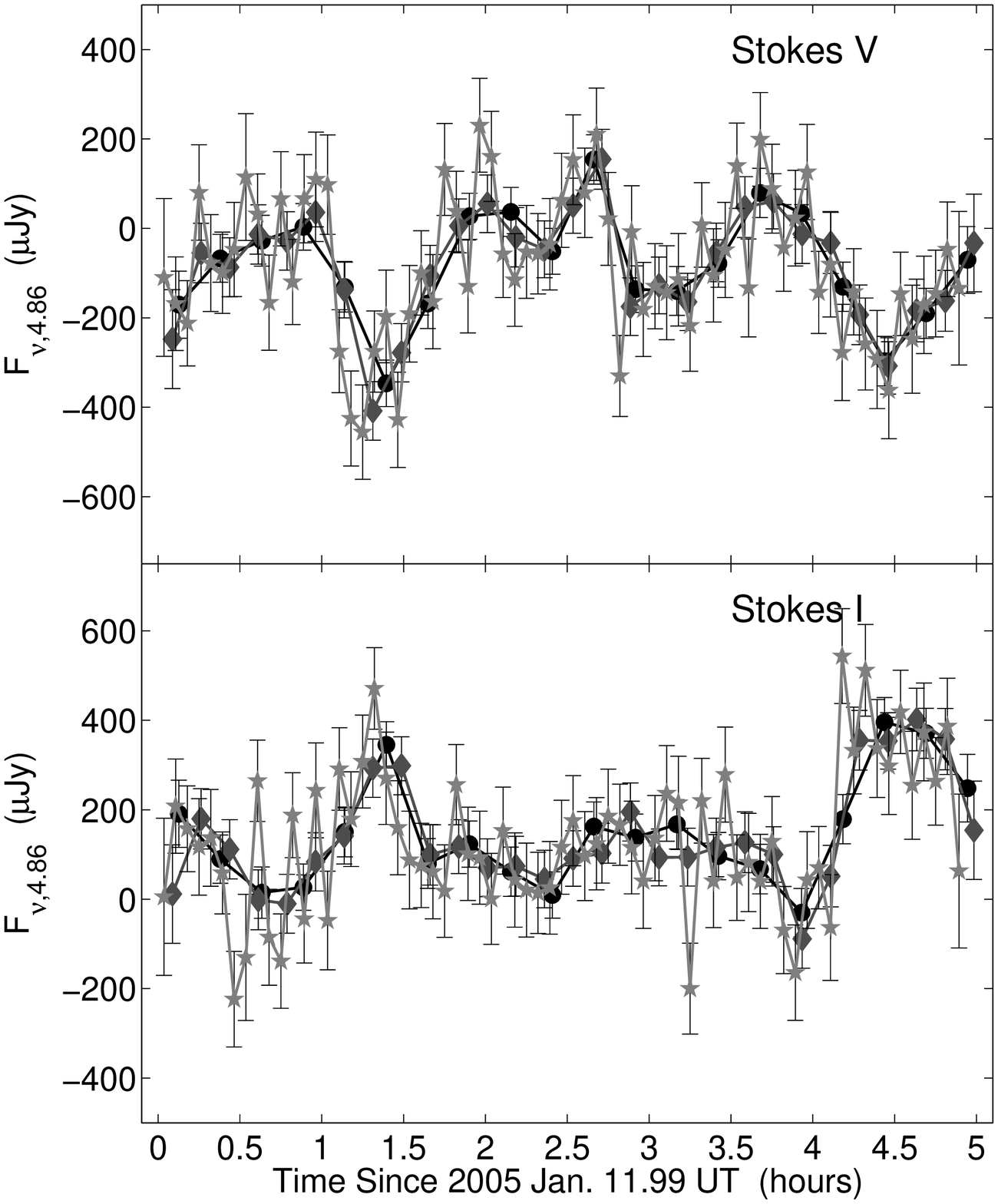,width=6.0in}}
\caption{Follow-up observations of \bd\ at 4.86 GHz taken on 2005, 
Jan.~11.99 UT for a total of 5 hours.  The 3-hour period detected in
the initial observations from Sep.~2002 is clearly seen.
\label{fig:0111}}
\end{figure}

To this point we have discussed the properties of the integrated radio
emission during the entire 7.8 hour simultaneous observation and the
follow-up observations.  However, from Figures~\ref{fig:xc},
\ref{fig:c}, \ref{fig:0110}, and \ref{fig:0111} it is clear that both
the total intensity and circularly polarized emission at 4.9 GHz are
strongly variable.  For the total intensity light curve the reduced
$\chi^2$ assuming a constant flux is $\approx 2.5-7.5$ for a time
resolution ranging from 5 min to 1 hour; for the circular polarization
light curve $\chi^2_r\approx 1.8-2.8$ over the same range of time
resolutions.  It is not clear if the variability of the weaker
emission at 8.5 GHz is significant, with $\chi^2_r\approx 1-1.7$
depending on the time resolution.

We are confident that this variability is not an observational
artifact for two reasons.  First, the light curves of two field
sources, constructed in the same manner as that of \bd\ (see
\S\ref{sec:rad}), do not exhibit significant variability, with
$\chi^2_r\approx 1.2$ and $\approx 1.8$ for a time resolution of $10$
min.  In fact, the reduced $\chi^2$ values for the field sources are
over-estimated compared to those for \bd\ since, unlike \bd, these
sources are spatially extended and positioned away from the phase
center.  As a result, changes in the synthesized beam during the
observation affect their light curves more significantly than that of
\bd.  Second, the variability of \bd\ is significant in circular
polarization where no field sources, which can introduce false
variability, produce detectable emission.  It is important to note
that in the two past observations of \bd, lasting 150 and 180 minutes,
the persistent emission was also variable (B02).

\begin{figure} 
\centerline{\psfig{file=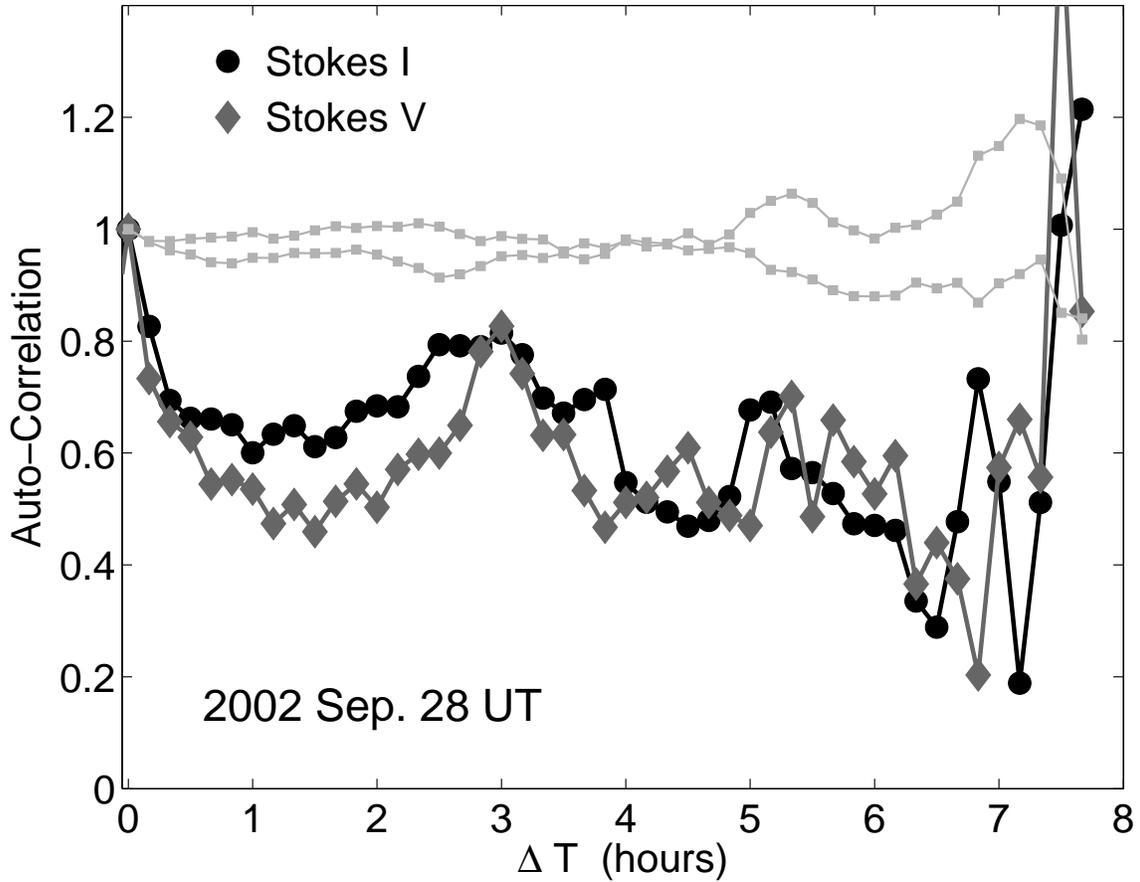,width=6.0in}}
\caption{Auto-correlation function (ACF) as a function of  temporal 
lag for the total intensity (black circles) and circular polarization
(gray diamond) light curves of \bdlong, as well as two field sources
(gray squares).  The time resolution is 10 minutes (thick lines and
field sources).  A clear peak in the ACF of \bd\ is seen at a lag of 3
hours.  The flat ACF of the field sources is indicative of a
non-periodic signal.  The timescale of about 1 hour on which the
signal from \bd\ de-correlates (i.e., the first minimum in the ACF)
indicates that the covering fraction of the emission region is $\sim
10\%$.
\label{fig:acf}}
\end{figure}

\begin{figure} 
\centerline{\psfig{file=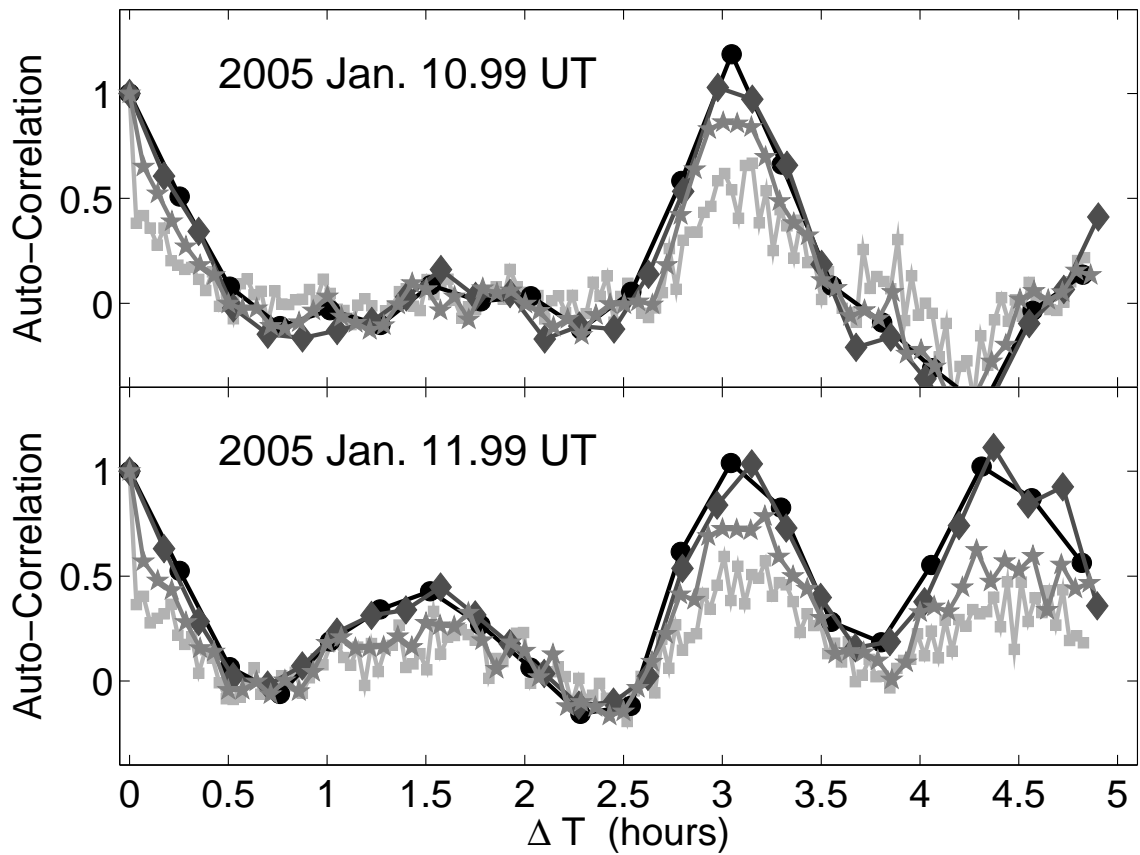,width=6.0in}}
\caption{Auto-correlation function (ACF) as a function of temporal 
lag for circular polarization in the follow-up observations on Jan.~10
(top) and Jan.~11 (bottom).  The time resolution ranges from 2 to 15
min.  A clear peak is seen at a lag of 3 hours, with a fall-off time
of about 1 hour, confirming the periodicity.
\label{fig:acf011011}}
\end{figure}

In addition to the variability, it appears from Figures~\ref{fig:c},
\ref{fig:0110}, and \ref{fig:0111} that the radio emission at 4.9 GHz
is periodic.  The auto-correlation function (ACF) of both the total
intensity and circular polarization at a wide range of time
resolutions shows a clear peak at $3$ hours (Figures~\ref{fig:acf} and
\ref{fig:acf011011}).  The ACFs of the two field sources described
above are flat, as expected for non-periodic sources.

We assess the statistical significance of the 3-hour periodicity using
two other methods which provide a measure of the power spectrum and
are therefore more easily calibrated: The Lomb-Scargle (LS)
periodogram (e.g., \citealt{ptv+92}), and a one-dimensional CLEAN
algorithm \citep{rld87}.  The latter method has the added advantage
that artifacts and spectral leakage which arise from the sampling
function and the finite length of the observation are removed.

\begin{figure} 
\centerline{\psfig{file=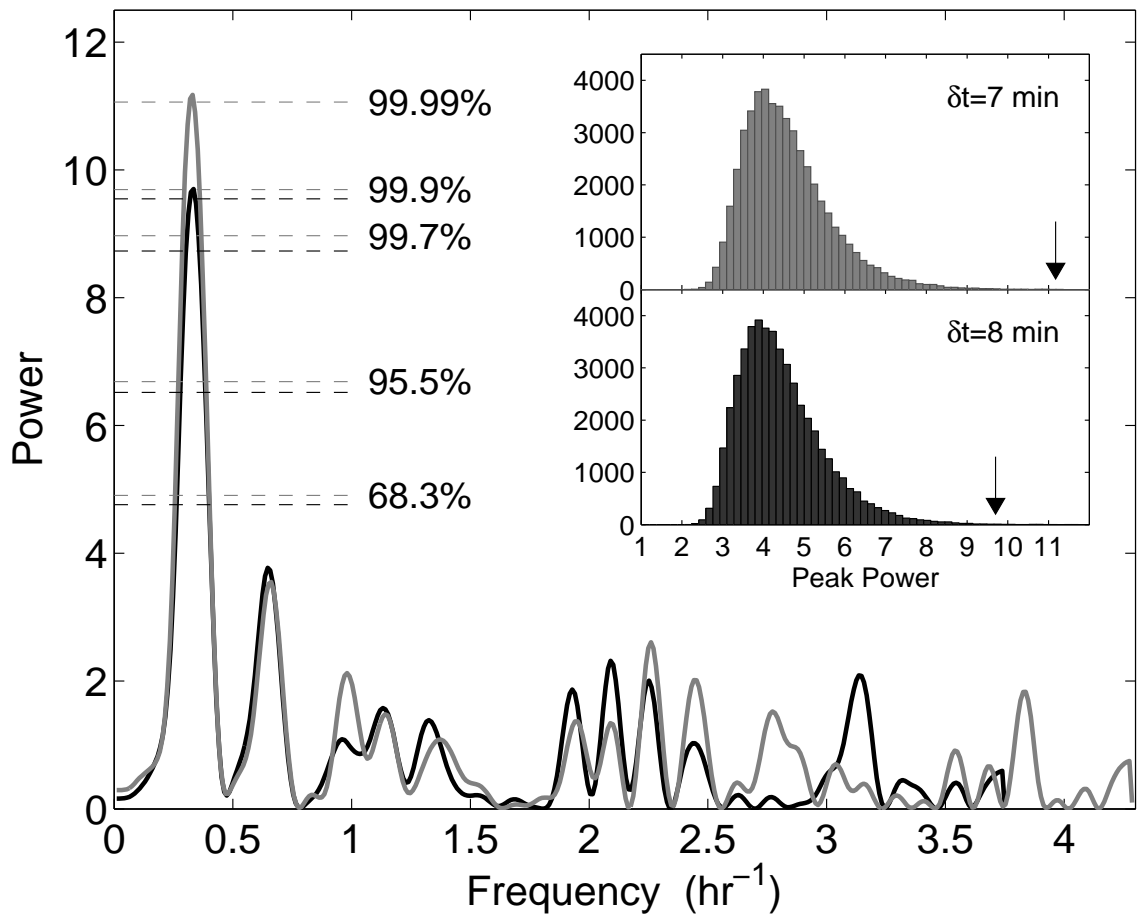,width=6.0in}}
\caption{Lomb-Scargle periodogram for the circularly polarized
emission at 4.9 GHz with a time resolution of 8 min (black) and 7 min
(gray).  The dashed lines indicate the $68.3\%$, $95.5\%$, $99.7\%$,
$99.9\%$, and $99.99\%$ significance levels based on Monte Carlo
simulations of $50,000$ randomized versions of the light curve
(inset).  For both time resolutions the most significant peak
($99.9\%$ and $99.99\%$) is at a frequency of 0.33 hr$^{-1}$, or a
period of 3.0 hours.  We find the same result, though with different
significance (see Figure~\ref{fig:ls_timeres}), for both total
intensity and circular polarization at a wide range of time
resolutions.
\label{fig:ls}}
\end{figure}

We assess the significance of the results using the Monte Carlo
method.  Specifically, we construct both randomized versions of the
4.9 GHz light curves, and simulated light curves with a ($\chi^2$)
variability similar to that of \bd.  Depending on the time resolution
we use $50,000$ to $100,000$ light curves in each set and perform the
same procedures as on the real data.  The significance of peaks in the
LS and CLEAN power spectra of the real light curve are determined as
the fraction of simulated light curves which produce stronger peaks.
We find that both the randomized and simulated sets provide a similar
measure of the significance, as do the LS and CLEAN procedures.  A
representative LS periodogram is shown in Figure~\ref{fig:ls}.

\begin{figure} 
\centerline{\psfig{file=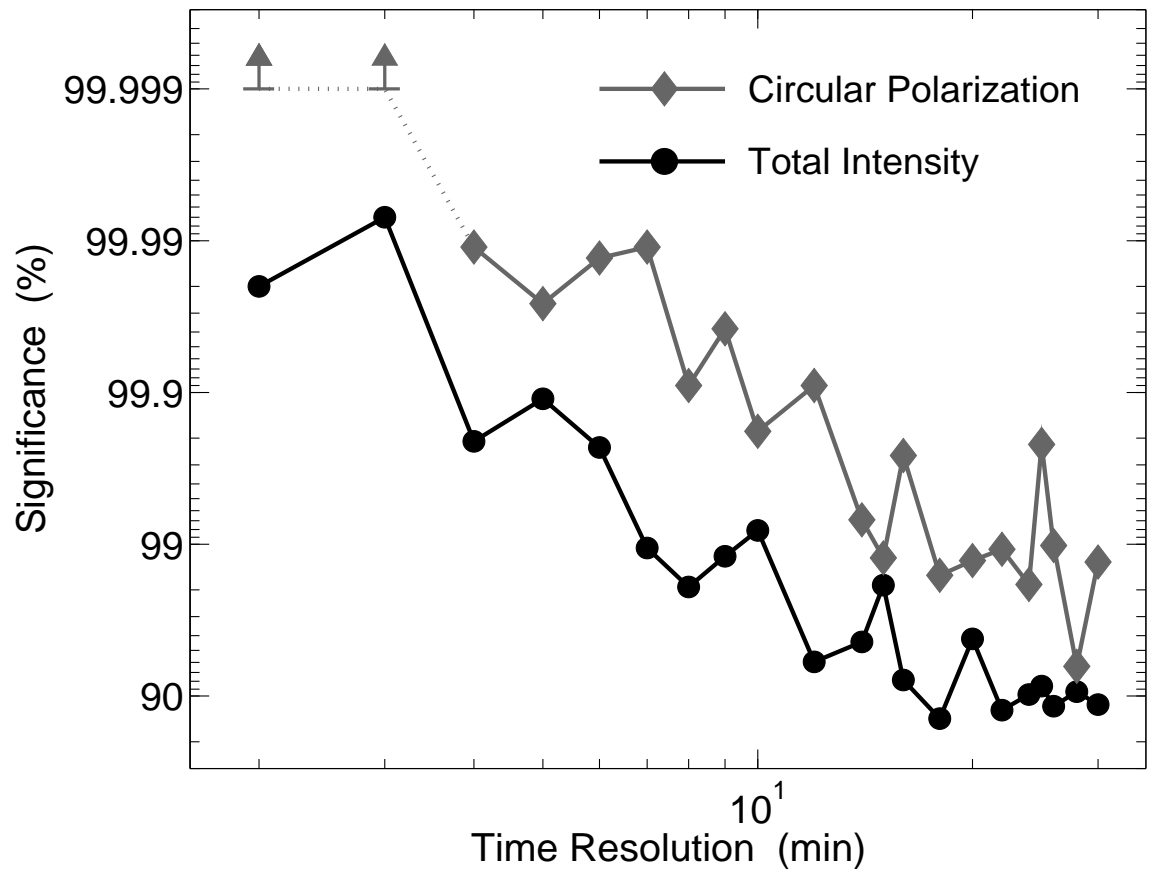,width=6.0in}}
\caption{Significance of the 3-hour period in the total intensity 
(black) and circular polarization (gray) 4.9 GHz light curves as a
function of time resolution.  The significance is determined using
Monte Carlo simulated and randomized light curves as described in
\S\ref{sec:period} and shown in Figure~\ref{fig:ls}.
\label{fig:ls_timeres}}
\end{figure}

\begin{figure} 
\centerline{\psfig{file=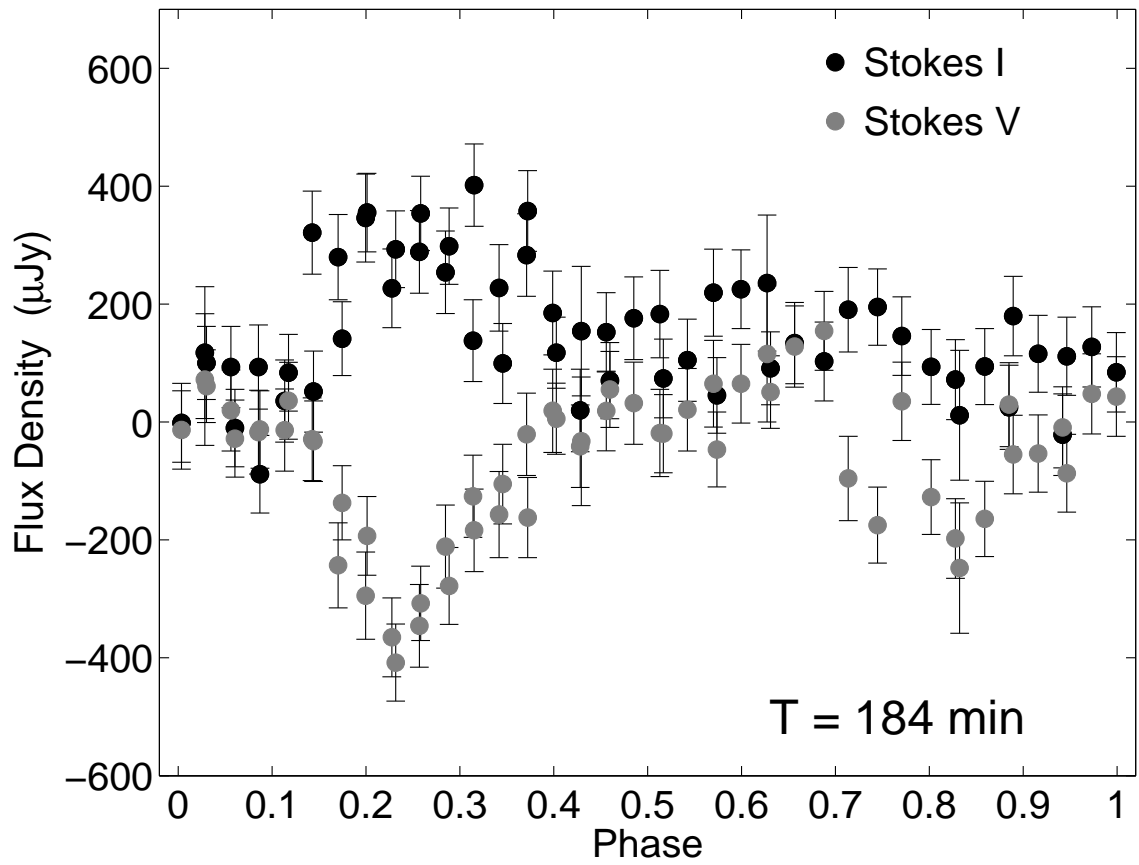,width=6.0in}}
\caption{Phased light curve (total intensity and circular 
polarization) for the observations from Jan.~10 and 11 UT folded with
a period of 184 min.
\label{fig:phase}}
\end{figure}

The significance of peaks in the LS and CLEAN power spectra does
depend on the choice of time resolution.  This is expected since as
the period is sampled more coarsely the strength of the signal is
expected to decrease.  We use time resolutions
ranging\footnotemark\footnotetext{We do not investigate time
resolutions smaller than 2 min since the computation time becomes
prohibitively long.} from 2 to 30 min (Figure~\ref{fig:ls_timeres}),
and find that the strongest peak always occurs at a frequency of
$0.33$ hr$^{-1}$, or a period of 3 hours.  For the simulated light
curves the peaks that exceed the value for \bd\ occur over the entire
frequency range.  We also find that the significance of the
periodicity in the circular polarization light curve is consistently
higher than for the total intensity light curve.  This is expected
because the contamination from field sources is non-existent in
circular polarization.  For the range of time resolutions investigated
we find that the significance of the 3-hour period peaks at $\approx
99.999\%$ for $\delta t=2$ min.  This result is confirmed by the
follow-up observations, and a phased light curve of the data from
2005, Jan.~10 and 11 UT, folded with a period of 184 min is shown in
Figure~\ref{fig:phase}.

Thus, the radio emission from \bd\ is periodic with $P=3$ hours.  This
period is maintained on timescales of at least two years.  If related
to the rotation of \bd\, then using $R_s\approx 6.3\times 10^9$ cm,
the inferred period indicates a surface equatorial rotation velocity
$v\approx 37$ km s$^{-1}$.  This value is nearly 2.5 times higher as
that inferred from high-resolution optical spectra, $v{\rm
sin}i\approx 15\pm 5$ km s$^{-1}$ \citep{sgh+01}, suggesting an
inclination angle, $i\approx 24\pm 8^\circ$.  Alternatively, if the
period is related to the orbital motion of a companion which induces
magnetic activity (see below) then the semi-major axis is about
$3.1\times 10^{10}$ cm or about five times the stellar radius; this is
similar to the case of the highly active RS CVn systems (e.g.,
\citealt{ml85}).  Finally, the periodic rise in flux may be a series
of weak flares, rather than variable persistent emission, in which
case the flare generation process is periodic.

\section{The Origin of the Magnetic Activity in \bd}
\label{sec:mag}

The results presented in the previous sections directly confirm for
the first time that the relative activity patterns in late-M and L
dwarfs differ from those in early-type stars and early M dwarfs.  It
is therefore crucial to address the origin of the magnetic fields in
the well-studied \bd\ as an example of a larger trend that is emerging
in sub-stellar objects.  While some dynamo models for convective stars
predict a reduced strength, physical scale, and lifetime of the
magnetic field (e.g., \citealt{ddr93}), the extent of this reduction
is still not known.  Clearly, the stability of the magnetic field of
\bd\ over a period of three years indicates a long-lived process.
Similarly, the physical scale of the field, $R\sim R_s$, is
significantly larger than the physical scale of the convection.
Finally, the inferred field strength and electron densities are
similar to those inferred in early M dwarfs.  It is therefore possible
that the current turbulent dynamo theories are missing some key
ingredients, or that the magnetic fields are generated and amplified
in another process, possibly by interaction with a close-in companion.

\subsection{The Influence of a Close-in Companion}
\label{sec:comp}

The enhancement of magnetic fields by a close companion may play a
role in the RS CVn class of active short-period binaries, and a
comparison is therefore particularly illustrative.  These systems are
known to produce radio gyrosynchrotron emission from magnetic fields
of about 200 G, with a typical luminosity of $L_{\nu,R}\approx 2\times
10^{16}$ erg s$^{-1}$ Hz$^{-1}$, a factor of $10^3$ higher than that
from \bd.  Using the scaling for gyrosynchrotron emission, $L_{\nu,R}
\propto B^{-3/4}R_s^2$, and taking into account the similar magnetic
field strengths and covering fractions, as well as the ratio of radii
($0.09$ R$_\odot$ vs.~$\sim 2$ R$_\odot$), we find $L_{\nu,{\rm 2M}}/
L_{\nu,{\rm RS\,CVn}}\approx 2\times 10^{-3}$, in close agreement to
the measured ratio.  In addition, the ratio $L_{\nu,R}/L_{\rm bol}$
for RS CVn systems is known to be correlated with the orbital period
of the system, such that tighter binaries are more radio active
\citep{ml85,dsl89}.  Extrapolating this relation to the value
$L_{\nu,R}/L_{\rm bol}\approx 6\times 10^{-17}$ Hz$^{-1}$ for \bd\ we
predict an orbital period of about 1 hour.  This is not so dissimilar
from the period of 3 hours inferred here.  Thus, it is possible that
\bd\ represents a low-mass scaled version of the RS CVn binaries and
that the same type of interaction that enhances the magnetic fields in
RS CVn systems may take place here.

Another illustrative example is the recent detection of enhanced
chromospheric Ca II H \& K emission from the star HD 179949 orbited by
a $0.84$ M$_J$ planet with an orbital period of about $3.09$ days, or
a semi-major axis of $6.7\times 10^{11}$ cm \citep{swb03}.  The
emission appears to be periodic and in phase with the orbit of the
planet.  The conclusion drawn by \citet{swb03} is that the planet
induces magnetic activity on the stellar surface which gives rise to
chromospheric emission.

These diverse examples illustrate that interaction with a companion
may enhance activity.  In this context both tidal and magnetic effects
can play a role \citep{csm00}.  In the case of tidal interaction, if
the orbital and rotation velocities are not the same, the generation
of time-varying tidal bulges may give rise to increased turbulent
motions or an enhanced $\alpha$-effect.  The synchronization timescale
(e.g., \citealt{gs66}) for a companion orbiting \bd\ is long, $t_{\rm
sync}\approx Q_s\omega (R_s^3/GM_s)(M_s/M_p)^2(a/R_s)^6\approx 1.6
M_{s,J}$ Gyr; a subscript $p$ designates the primary, a subscript $s$
designates the secondary, $\omega$ is the angular velocity, and $a$ is
the orbital separation.  Thus, it is likely that the putative
companion is not tidally locked and tidal interaction may be an
effective mechanism.

The strength of the tidal effect depends sensitively on both the
binary separation and the mass ratio, $\Delta g_p/g_p\propto (M_s/M_p)
[R_p/(a-R_p)]^3$.  This suggests that the tidal effect may be
particularly effective for late-M and L dwarfs where the primary mass
is low and the mass ratio may thus be of order unity.  For example,
for the HD 179949 system the distortion is $\Delta g_p/g_p\approx
1.3\times 10^{-6}$, while in the case of \bd\, assuming a Jupiter-mass
companion and using $a\approx 5R_p$ we find $\Delta g_p/g_p\approx 3
\times 10^{-4}$.

Direct magnetic interaction is also possible, with the magnetic energy
of the system proportional to the product of the primary and secondary
field strengths \citep{csm00}.  In this case the energy is $E_B\propto
B_pB_s/d_{m,p}^2d_{m,s}^2$, where $d_{m,s}$ is the magnetospheric
radius of the secondary and $d_{m,p}\equiv a-d_{m,s}$ is the distance
from the primary to that radius.  The energy flux is related to $E_B$
via $F\propto\epsilon E_Bv_c$, where $\epsilon$ is an efficiency
factor and $v_c$ is the combined velocity of the stellar magnetic
motions (or turbulence) and the relative velocity of the orbit (at the
stellar surface) compared to the rotation of the star.  Thus, as in
the case of tidal interaction, the possible larger mass ratio and
small separation in the case of \bd\ suggests that this mechanism may
be more efficient than in extra-solar planets orbiting solar-type
stars.

A final possibility is that a brown dwarf companion to \bd\ may
undergo sustained Roche lobe overflow which may result in enhanced
activity \citep{bkr+00,blk+02}.  This scenario has been proposed for
H$\alpha$ active dwarfs, although no direct evidence was found to
support this hypothesis.  Here we may for the first time have the
required evidence.  For sustained overflow to occur, the logarithmic
change in radius of the secondary, $d{\rm ln} R_L/d{\rm ln M_s}<-1/3$,
requires a mass ratio $q\equiv M_s/M_p <0.63$.  The period is given by
$P^2=4\pi^2a^3/GM_p(1+q)$ while the radius of the secondary is equal
to the Roche lobe radius, $R_s=R_L=0.49aq^{2/3}/[0.6q^{2/3}+{\rm
ln}(1+q^{1/3})]$.  The radius of the secondary is also given by
$R_s=(4/3)\pi(qM_p)^{-1/3}/[1+ 1.8(qM_p)^{-1/2}]^{4/3}$
\citep{zs69,ste91}, where the constant of $4/3$ is appropriate for an
age of $1-5$ Gyr \citep{blk+02}.  Using $P=3$ hr and $M_p=0.076$
M$_\odot$ we find $q\approx 0.14$, consistent with sustained mass
transfer, a secondary mass $M_s\approx 13$ M$_J$, a secondary radius
$R_s\approx 1.1$ R$_J$, and an orbital separation $a\approx 5.2R_p$.

\begin{figure} 
\centerline{\psfig{file=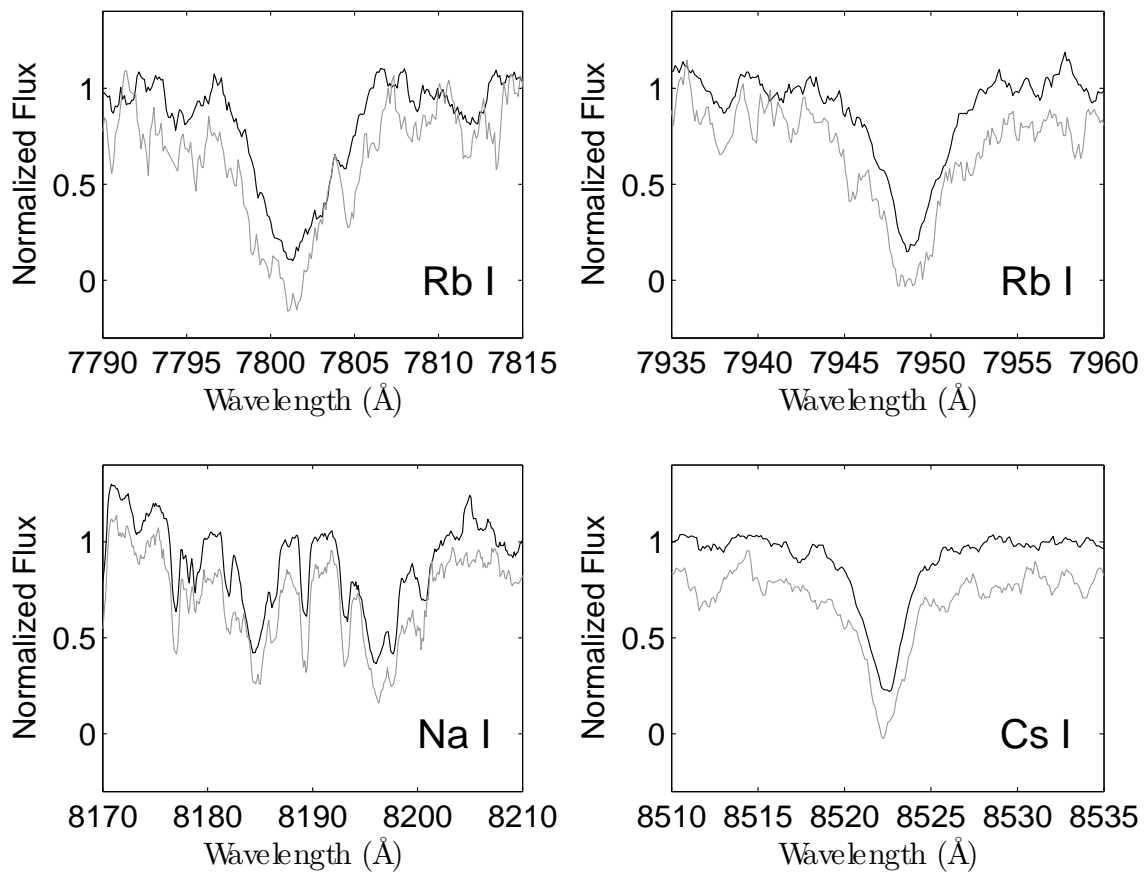,width=6.0in}}
\caption{Zoom-in on four spectral features in the high-resolution 
MIKE spectra of \bd, smoothed by two times the instrumental
resolution.  The second spectrum has been offset downward for clarity.
No obvious shift in the lines is discernible.
\label{fig:mike}}
\end{figure}

\begin{figure} 
\centerline{\psfig{file=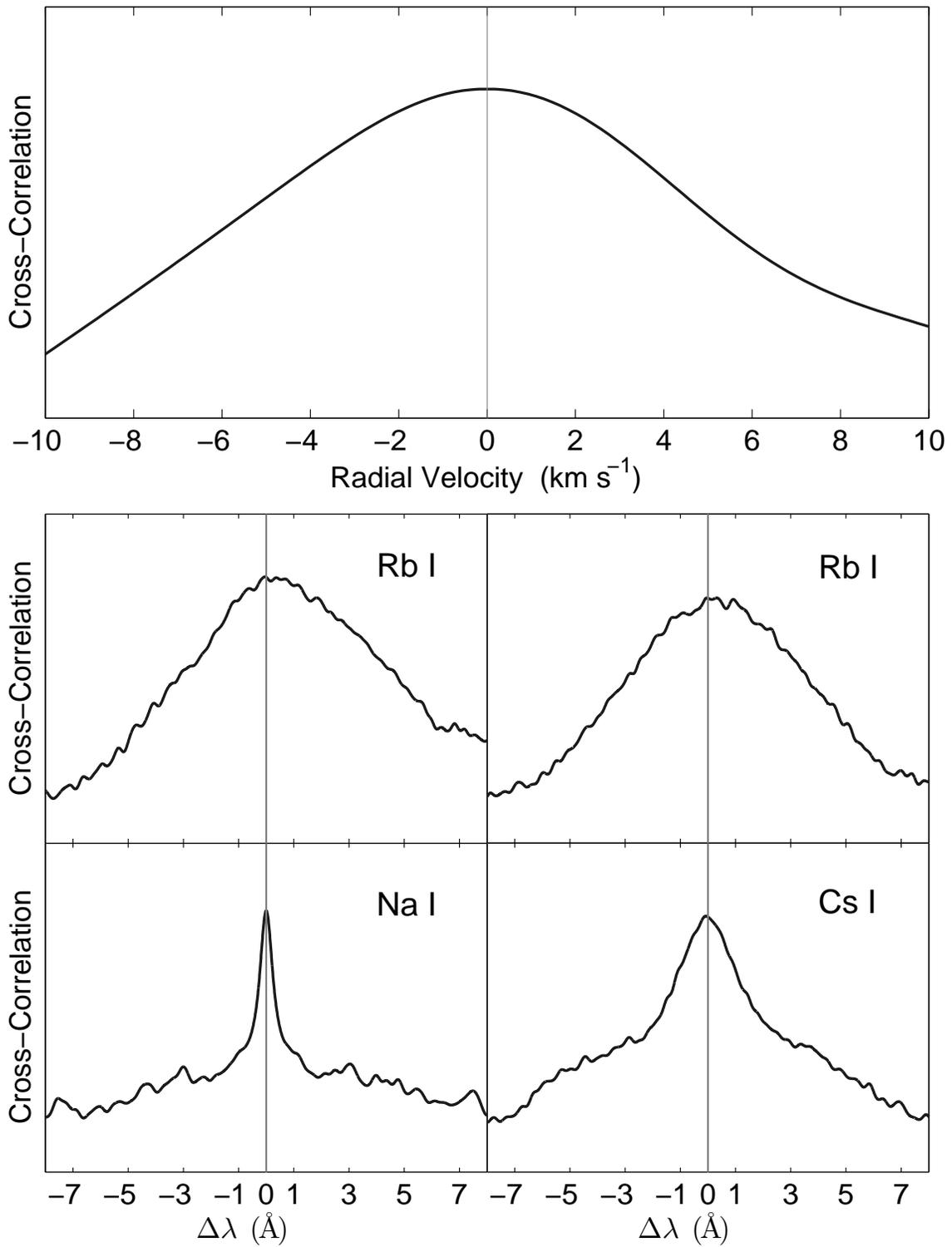,width=6.0in}}
\caption{Cross-correlation of the two MIKE spectra using the 
orders in which strong absorption lines were detected.  The peak is at
zero shift indicating no detectable radial velocity signature between
the two epochs.  The limiting factor in the accuracy of this result is
the width of the absorption features.
\label{fig:cross_corr}}
\end{figure}

Is there a way to directly test the hypothesis of a close-in
companion?  As in the case of planetary companions around solar-type
stars, radial velocity, astrometric shift, and photometric techniques
may provide an answer.  The expected radial velocity amplitude
(assuming zero eccentricity) in the case of \bd\ is $A\approx
2.3\times 10^3 (P/3\,{\rm hr})^{-1/3} (M_s{\rm sin}i_o/M_J)(M_p/0.076
M_\odot)^{2/3}$ m s$^{-1}$ for a Jupiter-mass companion.  Using a
cross-correlation of the four strongest and narrowest absorption
features of \bd\ (Figure~\ref{fig:mike}) we obtain a limit of $A<4$ km
s$^{-1}$ from the pair of MIKE spectra separated by about half a
period (Figure~\ref{fig:cross_corr}).  This indicates that $M_s{\rm
sin}i_o<1.7$ M$_J$ if the limit is on the maximum amplitude.
Naturally, if the two epochs were obtained when the putative companion
was located $\sim 90^\circ$ away from the line of sight to \bd, the
lack of a radial velocity signature does not provide a useful limit.

For an orbital inclination ${\rm cos}i_o<(R_p+R_s)/a$, or $i_o\gtrsim
68^\circ$, we expect to observe an eclipse as the companion occults
\bd\ ($i_o=0^\circ$ corresponds to a face-on orbit).  For $i_o\gtrsim
73^\circ$ more than half the surface of \bd\ would be occulted
resulting in large photometric variability.  If all inclinations are
equally likely, the probability of eclipse is about 24\%.
\citet{gmh+02} find no significant variability for \bd\ in their
I-band study of photometric variability in L dwarfs.  Observations
were obtained with a time separation as short as $1-2$ hours, while
the maximum time baseline investigated was 53 days.  The lack of
variability indicates that $i_o<68^\circ$ {\it if} \bd\ is in fact
orbited by a close companion.

Finally, a companion will induce an astrometric shift given by $\theta
\approx 0.1M_sM_p^{2/3}(P/3\,hr)^{2/3}$ mas for a $2:1$ mass ratio.
This is just within the reach of current Very Long Baseline
Interferometry (VLBI) observations, but lower mass ratios will result
in an undetectable shift.  In principle, an astrometric shift down to
a mass ratio of $50:1$, or $M_s\approx 1.7$ M$_J$, may be detected
with the 4 $\mu$as precision of the Space Interferometry Mission
(SIM).

\subsection{Magnetic Structure and Rotation}

The 3-hour period may alternatively be due to the rotation of \bd.
Using the inferred radius of 0.09 R$_\odot$, the equatorial velocity
is 37 km s$^{-1}$, leading to an inclination angle of $24\pm 8^\circ$,
or nearly pole-on.  The azimuthal orientation, $\phi$, of the axis of
rotation relative to the line of sight is not known.  However, given
the low inclination and the large covering fraction, the
radio-emitting region has to be located at a low latitude for most
$\phi$ values.  This is because a location near the pole would make
the bulk of the region visible throughout the rotation period,
resulting in a nearly constant flux level.  As can be seen from
Figure~\ref{fig:c} the flux drops to nearly zero between the peaks
indicating the emission region is fully occulted for about half the
rotation period.  We note also that the low inclination may explain
the constant sign of the circular polarization since for most values
of $\phi$ only one of the hemispheres is visible.

If the 3-hour period is related to the rotation of \bd\ then we can
estimate the Rossby number, $Ro=P/\tau_c$, which is relevant for
dynamo models.  The convective turnover time for \bd\ is $\tau_c=
(MR^2/L)^{1/3}\approx 0.8$ yr indicating that $Ro\approx 4.4\times
10^{-4}$.  At such low values early M dwarfs exhibit saturated X-ray
emission, $L_X/L_{\rm bol}\sim 10^{-3}$ \citep{pmm+03}.  Clearly, the
conditions in L dwarfs are sufficiently different that a low Rossby
number does not result in X-ray emission.  We also note that the large
value of $\tau_c$ may explain the stability of the radio emission
(flux and circular polarization) over a period of about three years.

\subsection{Periodic Flaring}

Finally, it is possible that the variable persistent emission is in
fact a series of periodic flares with a rate of occurrence of about
0.33 hr$^{-1}$.  The frequency of strong flares, similar to the one
detected by B01 with a flux of about 720 $\mu$Jy, is $\lesssim 0.04$
hr$^{-1}$.  This pattern is similar to the case of the M9 dwarf
LHS\,2065 for which strong H$\alpha$ flares have an occurrence rate of
$\lesssim 0.03$ hr$^{-1}$ while weak flares occur at a rate as high as
about 0.5 hr$^{-1}$ \citep{ma01}.  If this is the case here, then the
timescale to build up magnetic stresses is about 3 hours, while the
lifetime of the electrons, corresponding to the width of the flares,
is about 1 hour.

In this scenario it is possible that the periodic weak flares are
inefficient at heating the corona and chromosphere, thus explaining
the lack of accompanying X-ray and H$\alpha$ emission.  The more rare
strong flares, on the other hand, with an impulsive energy release a
few times larger may be accompanied by efficient heating.  However, we
note that the time-integrated energy release in the putative flares
observed here, $E\sim \nu L_\nu t\sim 4\times 10^{26}$ erg, is not so
different from the integrated energy release inferred for the stronger
flare (B02).

\section{Emerging Activity Patterns of late-M and L dwarfs}
\label{sec:trends}

The growing sample of late-M and L dwarfs observed in the radio and
X-rays allows to investigate the patterns of activity as a function of
relevant physical parameters.  H$\alpha$ observations of late-M and L
dwarfs indicate a significant drop in the level of emission compared
to the bolometric luminosity (Figure~\ref{fig:activity_sptype}).
Objects earlier than about M6 tend to have a ratio $L_{H\alpha}/L_{\rm
bol}$ in the range of $10^{-4}$ to $10^{-3}$.  However, by spectral
type L0 this ratio is typically lower than $10^{-5}$.  The few late-M
and L dwarfs which exhibit flares ($\sim 1\%$; \citealt{lkc+03})
appear to have levels of emission similar to those of early M dwarfs,
with only a single source exceeding that level to date.  Thus,
H$\alpha$ emission drops significantly with a decreased surface
temperature (later spectral type), and flares exceeding the saturation
value are extremely rare.

Similarly, the ratio $L_X/L_{\rm bol}$ drops from a saturation value
of about $10^{-2.5}$ in mid-M dwarfs to less than $10^{-4}$ for the
few late-M dwarfs observed to date (e.g.,
\citealt{fgs+93,rbm+00,mb02}).  With the exception of a
recently-discovered M9 dwarf for which a ratio of about 0.1 was
measured during a flare \citep{hss+04}, the flares observed in other
late-M dwarfs do not exceed the saturation level observed in the mid-M
dwarfs \citep{ste04}.  Thus, in both H$\alpha$ and X-rays the
saturation effect appears to be real, with flares replacing persistent
emission in late-M and L dwarfs.  \bd\ appears to follow the general
trend with upper limits on the X-ray and H$\alpha$ emission that are
about three orders of magnitude below the respective saturation
values.

On the other hand, the radio activity in late-M and L dwarfs, and in
particular \bd, exhibits a completely different trend.  The observed
ratio $L_R/L_{\rm bol}$ is about $10^{-8}$ for persistent emission
from early and mid-M dwarfs.  However, the level of radio emission
from late-M and L dwarfs exceeds this value by about an order of
magnitude, while flares are brighter by at least two orders of
magnitude.  \bd\ in particular has a ratio $L_R/L_{\rm bol}$ of about
$10^{-6.2}$.  Thus, while a significantly smaller number of objects
have been observed in the radio compared to H$\alpha$, {\it every}
single detection is brighter than the value observed in mid-M dwarfs.
Moreover, the fraction of detected objects is much higher, $\sim
40\%$.  This suggests that a saturation effect does not exists in the
radio band.  Moreover, the radio luminosity does not seem to decrease
with surface temperature.

Another important trend, proposed by B02, is that rapid rotators
exhibit stronger radio activity.  This is shown in
Figure~\ref{fig:activity_vel}.  In H$\alpha$ there is a clear
rotation-activity relation in dwarfs earlier than M7, but this
relation clearly breaks down in late-M and L dwarfs.  A similar
relation exists in X-rays, but it again breaks down in late-M and L
dwarfs.  In the radio band, on the other hand, the rotation-activity
relation appears to extend to late-M and L dwarfs.  The only objects
for which the ratio of radio to bolometric luminosity is lower than
the detected objects are those with $v{\rm sin}i\lesssim 10$ km
s$^{-1}$.

Thus, it appears that the efficiency of magnetic field generation and
dissipation in fact does not decrease in late-M and L dwarfs, but the
efficiency of chromospheric and coronal emission does.  Similarly, the
rotation activity relation appears to still hold, but the activity is
manifested in radio emission instead of H$\alpha$ or X-rays.  The
detection of rare H$\alpha$ and X-ray flares does suggest that
chromospheric and coronal heating may still occur, but this may
require large reconnection events, possibly accompanied by large radio
flares.

\section{Conclusions}
\label{sec:conc}

The simultaneous, multi-wavelength observations presented in this
paper provide unparalleled insight into the magnetic field properties
of an L dwarf.  Most importantly, these observations directly confirm
that radio activity is more prevalent in late-M and L dwarfs compared
to H$\alpha$ and X-ray activity.  The detection rate in the radio may
be as high as $40\%$, while in H$\alpha$ it is at most a few percent.
Only a few objects later than M7 have been observed in the X-rays, but
persistent emission is clearly rare.  A detailed analysis of the radio
emission along with the lack of detectable X-ray and H$\alpha$
emission give rise to the following inferences:

\begin{enumerate}

\item The radio emission from \bd, as compared to the bolometric
luminosity, is the strongest observed in any dwarf star to date.
Moreover, the emission from \bd\ violates the radio/X-ray correlation
by at least four orders of magnitude suggesting a drastic change in
the coronal and chromospheric conditions compared to those observed in
active stars down to spectral class M7.  This is the first direct
confirmation of this violation which has been observed in two
additional late-M dwarfs (B01; B02).

\item The radio emission is due to gyrosynchrotron radiation in a
magnetic field of about 175 G.  The size of the emission region is
about $10^{10}$ cm ($\sim 1.4R_s$), indicating a large covering
fraction at the surface.  These characteristics are similar to those
inferred in early M dwarfs.

\item The constant fraction and sense of the circular polarization in
observations conducted over a period of more than three years, implies
that an ordered magnetic field is maintained stably on long
timescales, despite the occasional production of strong flares (B02);
the incidence rate for the latter is $<0.04$ hr$^{-1}$.

\item The radio emission is periodic ($\gtrsim 99.999\%$
confidence level) with a period of $3$ hours.  If related to the
rotation period this indicates a rotation velocity of about 37 km
s$^{-1}$, while if related to an orbital period it indicates a
semi-major axis of about five times the stellar radius.
Alternatively, the period may be due to periodic release of the
magnetic energy in the form of weak radio flares.

\end{enumerate}

Clearly, the radio emission from \bd\ requires a large-scale,
long-lived, and relatively strong magnetic field.  The lack of
accompanying H$\alpha$ and X-ray emission suggests that the general
decline in these activity indicators in late-M and L dwarfs is not a
reliable tracer of decreased magnetic activity.  In particular, if the
magnetic field is essentially decoupled from gas in the cool
atmospheres of late-M and L dwarfs so that magnetic stresses are
suppressed, this may quench the chromospheric and coronal emission.
However, at larger distances the rate of collisions between charged
and neutral particles may drop and the field could then couple
effectively to the gas \citep{mbs+02}.  This is the region where radio
emission is expected to be generated most effectively, explaining the
strong radio signal compared to the reduced H$\alpha$ and X-ray
emission.  However, even this scenario cannot easily account for the
severe violation of the radio/X-ray correlation, or for the detection
of H$\alpha$ and X-ray flares from some L and T dwarfs.

While the origin of the observed periodicity remains unknown, all
three scenarios provide interesting constraints for turbulent dynamo
models.  In the case of periodicity due to a close-in companion, it is
possible that the magnetic field of \bd\ is amplified by interaction
with the companion.  This may alleviate the need to generate strong,
long-lived fields from convection alone.  If the incidence of activity
in the radio band is truly 40\%, this may also require an unusually
high fraction of companions, especially in comparison to current
estimates of the binary fraction in late-M and early L dwarfs of $\sim
15\%$ \citep{csf+03,grk+03}.  However, this fraction is estimated
primarily from direct detection with AO systems and the {\it Hubble
Space Telescope} which are only sensitive to companions at separations
of $\gtrsim 1$ AU, several hundred times larger than the possible
orbit in the case of \bd.  Direct evidence for close-in companions is
required before strong claims can be made, but radio active late-M and
L dwarfs may provide a signpost for binary systems.

On the other hand, in the case of periodicity due to rotation, it is
clear that any dynamo model has to explain the presence of magnetic
structures similar to those of M dwarfs without the benefit of a
radiative-convective transition zone and without hydrogen burning as a
source of heat.  Obviously, if the observed radio emission is in fact
due to a series of episodic flares, this places stringent constraints
on the timescale of magnetic field amplification and dissipation
($\sim 1$ hour) over a large fraction of the stellar surface.

The multi-wavelength approach taken in this paper is clearly the first
step in addressing the details of magnetic field generation at the
bottom of the main sequence and beyond.  Continued studies,
particularly in the radio and X-ray poorly-sampled range M9 to L5
(Figure~\ref{fig:activity_sptype}), are essential for understanding
individual objects, as well as the effects of various physical
properties (e.g., binarity, rotation, temperature, age).  These
observations will for the first time provide robust constraints for
dynamo models and for the structure of late-M and L dwarfs all the way
from the interior to the corona.

\acknowledgements We thank first and foremost Barry Clark (NRAO),
Scott Wolk (CXC, HEAD), and Meghan McGarry (CXC, HEAD) for their
efforts in scheduling the simultaneous VLA and Chandra observations.
Support for this work was provided by NASA through Chandra Award
Number GO2-3012B issued by the Chandra X-Ray Observatory center, which
is operated by SAO for and on behalf of NASA under contract
NAS8-39073.  Further support was provided by National Science
Foundation grant NSF PHY 99-07949.  EB is supported by NASA through
Hubble Fellowship grant HST-01171.01 awarded by the Space Telescope
Science Institute, which is operated by AURA, Inc., for NASA under
constract NAS 5-26555.

\begin{appendix}
\section{X-ray Observations of the L5 Dwarf 2MASS J15074769-1627386}
\label{sec:2m1507}

In addition to the X-ray observations of \bd\ outlined in this paper,
we also observed the L5 dwarf 2MASS J15074769-1627386 (hereafter,
2M1507-16; \citealt{rkg+00,krl+00}) as part of our {\it Chandra} AO3
program.  This object is located at a distance of $7.4$ pc
\citep{rkg+00,rgk+01,dhv+02}, has a surface temperature of about 1630
K, a bolometric luminosity of $10^{-4.27}$ L$_\odot$, an inferred
radius of 0.09 R$_\odot$ \citep{vhl+04}, and a rotation velocity
$v{\rm sin}i\approx 27$ km s$^{-1}$ \citep{bai04}.  The upper limit on
H$\alpha$ emission is $L_{H\alpha}/L_{\rm bol}<10^{-5.76}$
\citep{rkg+00} and no Li is detected to a limit of 0.1\AA.  Finally,
B02 find an upper limit on the radio luminosity at 8.5 GHz of
$3.8\times 10^{12}$ erg s$^{-1}$ Hz$^{-1}$ from a 2.2-hr observation,
corresponding to $L_R/L_{\rm bol}\lesssim 10^{-6.8}$.

The details of the X-ray observation and data analysis are given in
\S\ref{sec:xray}.  No X-ray source is detected at the position of
2M1507-16.  The number of counts in the low and total energy bands
were 2 and 4 respectively, while we expect 2 and 5 counts from
background alone.  We find no evidence for variability: In the total
energy band, the shortest time-span which includes 3 counts is 2330 s,
which we find from a Monte Carlo simulation, will occur in 22\% of
27.7 ks light curves containing four counts.  The 90\% confidence
limit on the time-averaged flux ($<$7 counts per 27.7 ks) is $<1\times
10^{-15}$ erg cm$^{-2}$ s$^{-1}$.  Taking the 3 counts within 2330 s,
we place a 90\% confidence upper limit on a 1-hr peak flaring flux of
$<1\times 10^{-14}$ erg cm$^{-2}$ s$^{-1}$.  At the distance of
2M1507-16, the upper limit on persistent emission is $L_X\lesssim
6.6\times 10^{24}$ erg s$^{-1}$, or ${\rm log}(L_X/L_{\rm bol})
\lesssim -4.5$.

\begin{figure} 
\centerline{\psfig{file=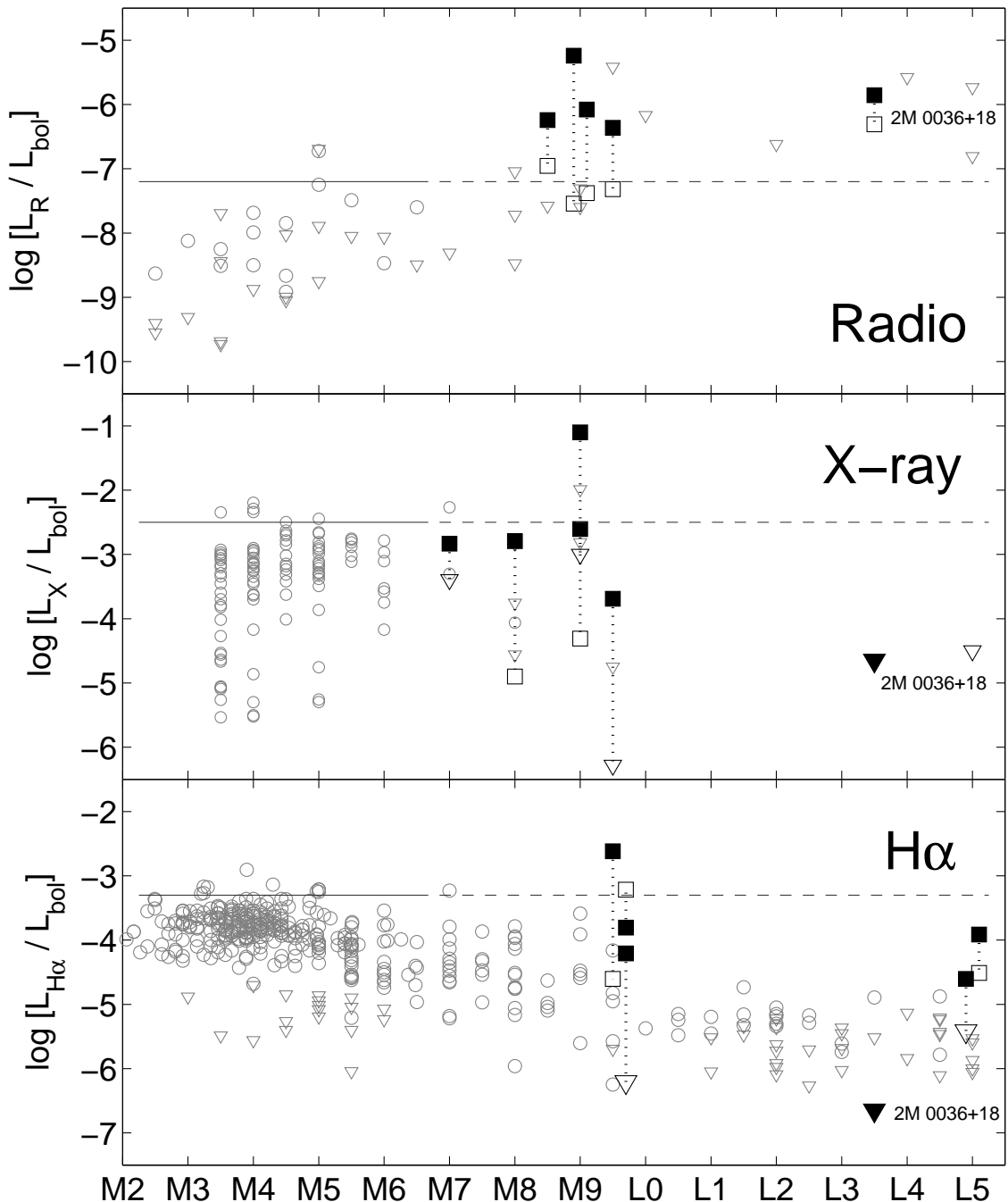,width=6.0in}}
\caption{Radio (top), X-ray (middle) and H$\alpha$ (bottom) activity 
as a function of spectral type.  Shown are objects with flares (solid
squares) and persistent emission (detections: open squares; upper
limits: triangles), as well as objects with persistent emission only
(detections: circles; upper limits: triangles).  The solid line in
each panel marks the maximum level of activity for spectral types
earlier than M7, and the dashed lines are an extrapolation to later
spectral types.  In both X-rays and H$\alpha$ the level of persistent
activity drops significantly beyond M7, while flares occasionally reach 
the same level of activity of early M dwarfs.  In the radio band, on 
the other hand, several objects exhibit activity levels much stronger 
than that observed in early M dwarfs.  This is particularly pronounced 
in the case of \bd.
\label{fig:activity_sptype}}
\end{figure}

\begin{figure} 
\centerline{\psfig{file=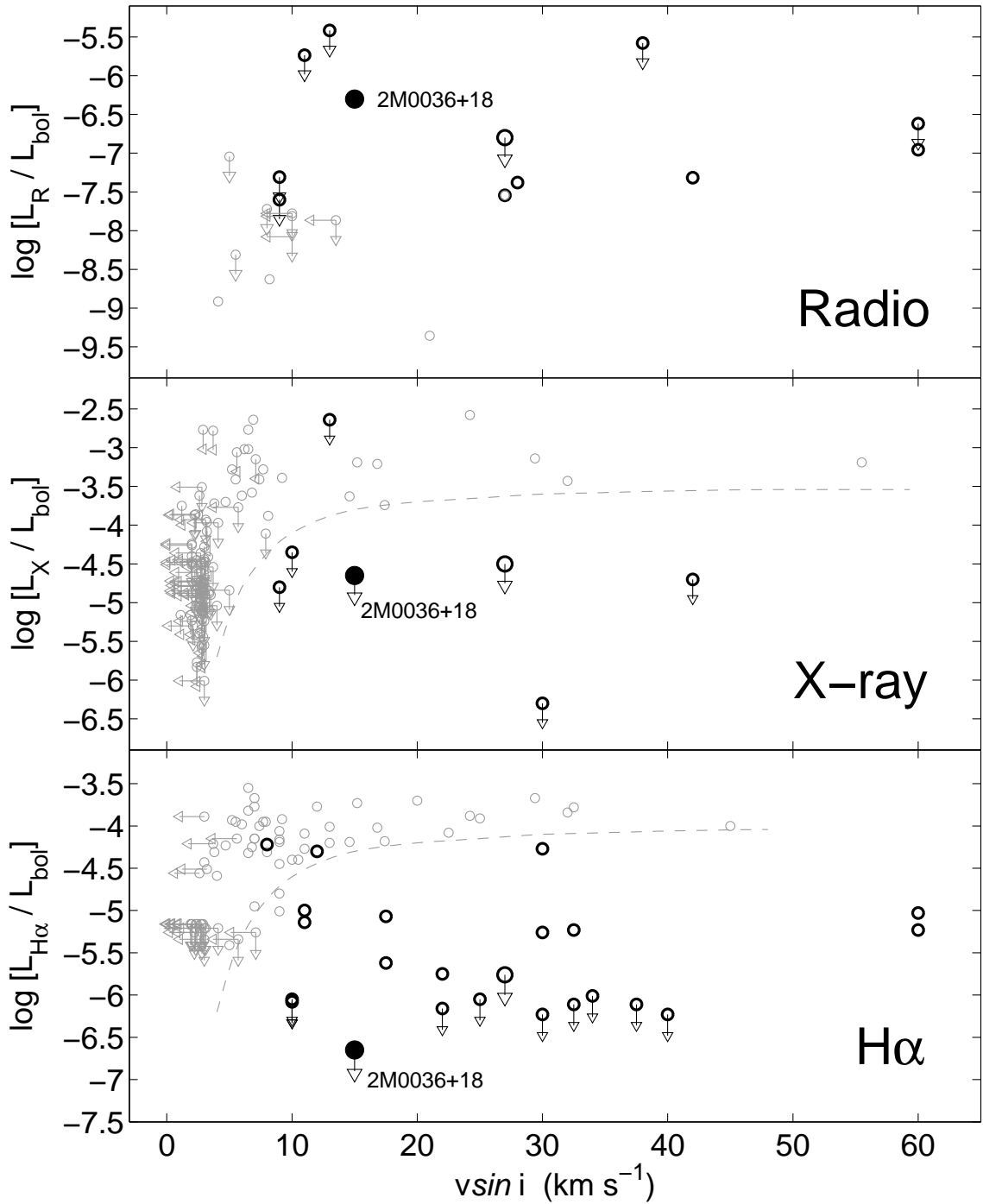,width=6.0in}}
\caption{Radio (top), X-ray (middle) and H$\alpha$ (bottom) activity 
as a function of rotation velocity.  M dwarfs up to spectral type M8
are shown in gray symbols, while dwarfs later than M8 are shown in
black symbols.  The early M dwarfs exhibit a clear rotation-activity
relation in H$\alpha$ and X-rays (gray dashed lines).  This relation
breaks down in the late-M and L dwarfs with many fast rotators
exhibiting no H$\alpha$ or X-ray emission.  In the radio, on the other
hand, faster rotation appears to be correlated with activity even for
dwarfs later than M9.  The large open circle designates the L5 dwarf
2M\,1507-16 (see Appendix).
\label{fig:activity_vel}}
\end{figure}

The upper limits on X-ray, radio, and H$\alpha$ activity for 2M1507-16
are shown in Figures~\ref{fig:activity_sptype} and
\ref{fig:activity_vel}.  Clearly this object follows the general trend
of decreased H$\alpha$ and X-ray activity in L dwarfs.  The radio
activity is at least a factor of four lower than that of \bd, but the
upper limit is several times brighter than other late-M and L dwarfs
with a similar rotation velocity.

\end{appendix}


\clearpage

\clearpage

\clearpage

\clearpage

\clearpage

\clearpage

\clearpage

\clearpage

\clearpage

\clearpage

\clearpage

\clearpage

\clearpage

\clearpage

\clearpage

\end{document}